\documentclass[aps,prc,twocolumn,showpacs,showkeys,nofootinbib,superscriptaddress]{revtex4-1}

\usepackage{dcolumn}
\usepackage{bm}
\usepackage{amsmath}    
\usepackage{amsfonts} 
\usepackage{amssymb}
\usepackage{mathptmx}      
\usepackage{graphicx}   
\usepackage{mathtools}
\usepackage{graphics}
\usepackage{multirow}
\usepackage{bibentry}
\usepackage{braket}
\usepackage{algorithm}
\usepackage{algorithmic}
\usepackage{listings}
\usepackage{setspace}
\usepackage{slashed}
\usepackage{color}
\usepackage{soul}

\usepackage{xfrac}

\def\al{\alpha}

\def\be{\begin{equation}}
\def\ee{\end{equation}}
\def\bea{\begin{eqnarray}}
\def\eea{\end{eqnarray}}

\usepackage{hyperref}
\hypersetup{colorlinks=true,citecolor=blue,linkcolor=blue,urlcolor=blue}

\makeatletter
\providecommand \@ifxundefined [1]{%
 \@ifx{#1\undefined}
}%
\providecommand \@ifnum [1]{%
 \ifnum #1\expandafter \@firstoftwo
 \else \expandafter \@secondoftwo
 \fi
}%
\providecommand \@ifx [1]{%
 \ifx #1\expandafter \@firstoftwo
 \else \expandafter \@secondoftwo
 \fi
}%
\providecommand \href@noop [0]{\@secondoftwo}%
\providecommand \href [0]{\begingroup \@sanitize@url \@href}%
\providecommand \@href[1]{\@@startlink{#1}\@@href}%
\providecommand \@@href[1]{\endgroup#1\@@endlink}%
\providecommand \@sanitize@url [0]{\catcode `\\12\catcode `\$12\catcode
  `\&12\catcode `\#12\catcode `\^12\catcode `\_12\catcode `\%12\relax}%
\providecommand \@@startlink[1]{}%
\providecommand \@@endlink[0]{}%
\providecommand \url  [0]{\begingroup\@sanitize@url \@url }%
\providecommand \@url [1]{\endgroup\@href {#1}{\urlprefix }}%
\providecommand \urlprefix  [0]{URL }%
\providecommand \selectlanguage [0]{\@gobble}%
\providecommand \bibinfo  [0]{\@secondoftwo}%
\providecommand \bibfield  [0]{\@secondoftwo}%
\providecommand \BibitemShut  [1]{\csname bibitem#1\endcsname}%
\let\auto@bib@innerbib\@empty

\begin{document}
\title{\boldmath
Light-front dynamic analysis of  the longitudinal charge density using the solvable scalar field model
in (1+1) dimensions}
\author{Yongwoo Choi}
\affiliation{Department of Physics, Kyungpook National University, Daegu 41566, Korea}
     
\author{Ho-Meoyng Choi}
\email{homyoung@knu.ac.kr}
\affiliation{Department of Physics, Teachers College, Kyungpook National University, Daegu 41566, Korea}
     
\author{Chueng-Ryong Ji}
\email{crji@ncsu.edu}
\affiliation{Department of Physics, North Carolina State University,
Raleigh, NC 27695-8202} 
     
\author{Yongseok Oh}
\email{yohphy@knu.ac.kr}
\affiliation{Department of Physics, Kyungpook National University, Daegu 41566, Korea}
\affiliation{Asia Pacific Center for Theoretical Physics, Pohang, Gyeongbuk 37673, Korea}

\begin{abstract}
We investigate the electromagnetic form factor $F(q^2)$ of the meson by using the solvable $\phi^{3}$ scalar field model in $(1+1)$ dimensions. 
As the transverse rotations are absent in $(1+1)$ dimensions, the advantage of the light-front dynamics (LFD) with the light-front time $x^+ = x^0 + x^3$ 
as the evolution parameter is maximized in contrast to the usual instant form dynamics (IFD) with the ordinary time $x^0$ as the evolution parameter. 
In LFD, the individual $x^+$-ordered amplitudes contributing to $F(q^2)$ are invariant under the boost, i.e., frame-independent, while the individual 
$x^0$-ordered amplitudes in IFD are not invariant under the boost but dependent on the reference frame. 
The LFD allows to get the analytic result for the one-loop triangle diagram which covers not only the spacelike ($q^{2}<0$) but also timelike region ($q^{2}>0$). 
Using the analytic results, we verify that the real and imaginary parts of the form factor satisfy the dispersion relations in the entire $q^{2}$ space. 
Comparing with the results in $(3+1)$ dimensions, we discuss the transverse momentum effects on $F(q^2)$ . 
We also discuss the longitudinal charge density in terms of the boost invariant variable $\tilde z = p^+ x^-$ in LFD.
\end{abstract}
\maketitle

\section{Introduction}
\label{sec:1}

The formulation of light-front dynamics (LFD)  based on the equal light-front time $x^+ = x^0 + x^3$ quantization has shown 
remarkable advantages for calculations in elementary particle physics, nuclear physics, and hadron physics.  
In particular, the light-front (LF) formulation is an essential theoretical tool for the 3-dimensional imaging and 
femtography efforts in the 12~GeV upgraded Thomas Jefferson National Accelerator Facility (JLab) and in the future 
Electron-Ion Collider project, with the investigation of the form factors, the generalized parton distributions (GPDs), 
the transverse momentum distributions of hadrons, etc. 
Taking advantage of the LFD, one of the new experiments planned at JLab is to measure the transverse charge densities 
of hadrons~\cite{CHM14}, which are defined by the 2-dimensional Fourier transforms of the electromagnetic (EM) form factors 
describing the distribution of charge and magnetization in the plane perpendicular to the direction of a fast moving hadron~\cite{Soper77}.  
Due to the Lorentz invariance of the transverse distance and momentum under the longitudinal boost, the relativistically 
invariant analysis of the transverse charge density can be straightforwardly attained in the $(3+1)$ dimensional LFD.  
The transverse charge densities are also related to the GPDs~\cite{Burkardt00,Burkardt02,Diehl02} and their properties have been 
explored in a number of 
works~\cite{Miller07,Miller09,Miller09a,CV07,SW10,MSW10,VAMZ10,Miller18,Cedric}.
In particular, it was demonstrated that the transverse charge density defined by the two-dimensional Fourier transform 
can be obtained from the so-called ``Drell-Yan-West (DYW)" frame ($q^+=0$ and $q^2=-{\bf q}^2_\perp=-Q^2 <0$) in LFD 
using the scalar $\phi^3$ model in $(3+1)$ dimensions~\cite{Miller09}.
Although its utility is limited only to the spacelike region ($q^2 <0$) due to the intrinsic kinematic constraint $q^+=0$, 
the DYW formulation~\cite{BRS73,Weinberg66a,CM69,Sawicki91,Sa92} 
may be regarded as the most rigorous and well-established 
framework to compute the exclusive processes since it involves typically the particle number conserving valence contribution. 
Various studies of two-body bound-states in the $(3+1)$ dimensional LFD can also be found in the framework of
scalar~\cite{FFT73,Karmanov80,Karmanov81,Mueller83,BJS85,Sawicki85,Sawicki86,CJS87,BJ00}
and fermion field~\cite{BJ00,BCJ00,DNF97,DSFS98} models.

On the other hand, the LFD analysis of the longitudinal charge density is not as straightforward as in the analysis of 
the transverse charge density due to the nontrivial spacetime mixture of the LF spatial distance $x^- = x^0 - x^3$ as well as 
its conjugate momentum $p^+ = p^0 + p^3$.
It is noteworthy that the new variable $\tilde z = p^+ x^-$ was recently introduced~\cite{MB19} for the boost invariant analysis 
in the longitudinal direction.   
Although the same level of significant progresses as in the case of the transverse charge density is yet to be expected in the 
analysis of the longitudinal charge density, it may be worthwhile to facilitate the scalar $\phi^3$ model in the $(1+1)$ dimensional LFD
extending the previous LFD analyses in $(1+1)$ 
dimensions~\cite{SM88,MS89,GS90} 
restricted only for the spacelike momentum transfer 
region now to the entire kinematic regions including the timelike momentum transfers as well.
We note that the advantage of LFD is indeed maximized in $(1+1)$ dimensions due to the absence of the transverse rotations 
which are not kinematical but dynamical in LFD. 
As evidenced in solving the $(1+1)$ dimensional QCD with large $N_c$ limit~\cite{tHooft74b}, the solution was provided even 
analytically in LFD first~\cite{tHooft74b,Einhorn76} well before its canonical formulation~\cite{BG78} was presented in the instant 
form dynamics (IFD) based on the equal time $x^0$ quantization and later numerically solved in IFD~\cite{LWB87,JLLX17,JLXY18}. 
While in IFD the individual $x^0$-ordered amplitudes contributing to the form factor $F(Q^2)$ are not invariant under the boost, i.e., 
dependent on the reference frame, the advantage of the LFD with the LF time $x^+$ as the evolution parameter is maximized
due to the frame-independence or the boost invariance of the individual $x^+$-ordered amplitudes contributing to $F(Q^2)$.

The dramatic difference of LFD analysis of form factors in $(1+1)$ dimensions compared to the case of $(3+1)$ dimensions may also 
be attributed to the fact that the DYW frame cannot be taken as it is restricted to $q^2=0$ in $(1+1)$ dimensions. 
As the $q^+ \neq 0$ frame must be used in $(1+1)$ dimensions for $Q^2 \neq 0$, it is inevitable to encounter the non-valence diagram arising from the 
particle-antiparticle pair creation (the so-called ``Z-graph").
That is, both valence and non-valence contributions should be included simultaneously for the form factor analysis in $(1+1)$ dimensions. 
As mentioned earlier, the LFD analyses of scalar $\phi^3$ model in $(1+1)$ dimensions were reported 
in Refs.~\cite{SM88,MS89,GS90} 
which were though restricted only for the spacelike momentum transfer region. 
Once the $q^+ \neq 0$ frame is chosen, however, one does not need to restrict the analysis only for the spacelike region.

We may also compare the $(1+1)$ dimensional results with the previous $(3+1)$ dimensional results~\cite{CJ99} within the same 
solvable scalar $\phi^3$ model since it was shown numerically that the $(3+1)$ dimensional results analytically continued from 
the spacelike region $(q^2<0)$ coincide exactly with the results directly obtained in the timelike region $(q^2>0)$. 
Stemming from the detailed analysis of the solvable and manifestly covariant model within 
the framework of the $(3+1)$ dimensional LF calculations, we have also developed a new method to explore 
the timelike region directly in the $q^+ \neq 0$ frame for the transition form factor $F_{\mathcal{M}\gamma}(q^2)$ in 
the meson-photon transition process, $\mathcal{M}(p) \to \gamma^*(q) + \gamma(p')$~\cite{CRJ17}. 
Our direct calculation in the timelike region showed the complete agreement not only with the analytic continuation result from the spacelike 
region but also with the result from the dispersion relation (DR) between the real and imaginary parts of the form factor~\cite{CRJ17}.
This direct method of analyzing the timelike region appears to advance our previous analysis of a solvable model in $(3+1)$ dimensions 
for the phenomenologically more realistic LF quark model (LFQM).

In this work, we present the $(1+1)$-dimensional analysis of the form factor in the solvable model both for the spacelike 
region and the timelike region obtaining the analytic results both for the valence and non-valence contributions.  
Our model is essentially the (1 + 1)-dimensional quantum field theory model of Mankiewicz and Sawicki~\cite{SM88,MS89},
 which was reinvestigated by several others. (See, for example, Refs.~\cite{Sawicki91,Sa92,GS90,BH98,CJZero,SB1,SB2}.) 
In this model, the wave function is obtained as the solution of the covariant Bethe–Salpeter (BS) equation 
 in the ladder approximation with a relativistic version of the contact interactions~\cite{Sawicki91,Sa92}. 
 The covariant model wave function is a product of two free single-particle propagators, the Dirac delta function for the overall momentum
 conservation, and a constant vertex function. Consequently, all our form factor calculations show various ways of evaluating the 
 Feynman triangle diagrams in scalar field theory.
Previous results reported in Ref.~\cite{GS90} in the spacelike region were confirmed, but now the results are extended to 
the timelike region.   
In particular, the anomalous threshold is observed in the timelike region as in the case of $(3+1)$-dimensional analysis.
Apparent satisfaction of DR is shown explicitly and analytically. 
Longitudinal charge density is clearly identified with respect to ``intrinsic" versus ``apparent" charge densities. 
We also discuss the LF longitudinal charge density in terms of the newly introduced boost invariant variable $\tilde z$
of Ref.~\cite{MB19}.

This paper is organized as follows.
In Sec.~\ref{sec:2}, we derive the analytic forms for both spacelike and timelike EM form factors using the scalar $\phi^3$ model 
in $(1+1)$ dimensions.  
We obtain the explicit form of the imaginary part of the form factor in the timelike region so that the DR relation between
the real and imaginary parts of the form factor can be tested.
In Sec.~\ref{sec:3}, we discuss the difference between the intrinsic longitudinal charge density obtained from the Fourier transform 
of the form factor and the apparent charge density including the relativistic corrections such as the Lorentz contraction 
in the so-called ``Breit" frame.
The explicit form of mean-square charge radius in the longitudinal direction is also derived from the slope of the charge form factor.
We then discuss the LF longitudinal charge density in terms of the boost invariant
variable $\tilde z$.
Section~\ref{sec:4} presents our numerical results for the intrinsic longitudinal charge densities for scalar $(\pi, K, D)$ mesons and 
their EM form factors in both spacelike and timelike regions comparing them with the previous $(3+1)$ dimensional results of Ref.~\cite{CJ99}. 
We summarize and conclude in  Sec.~\ref{sec:5}. 
The explicit analytic forms of the valence and non-valence contributions to the form factor are presented in Appendix.


\section{\boldmath Form Factor for scalar $\phi^3$ model in (1+1) dimensions}
\label{sec:2}
\subsection{Form factor in spacelike region}

\begin{figure*}
\includegraphics[width=0.75\textwidth]{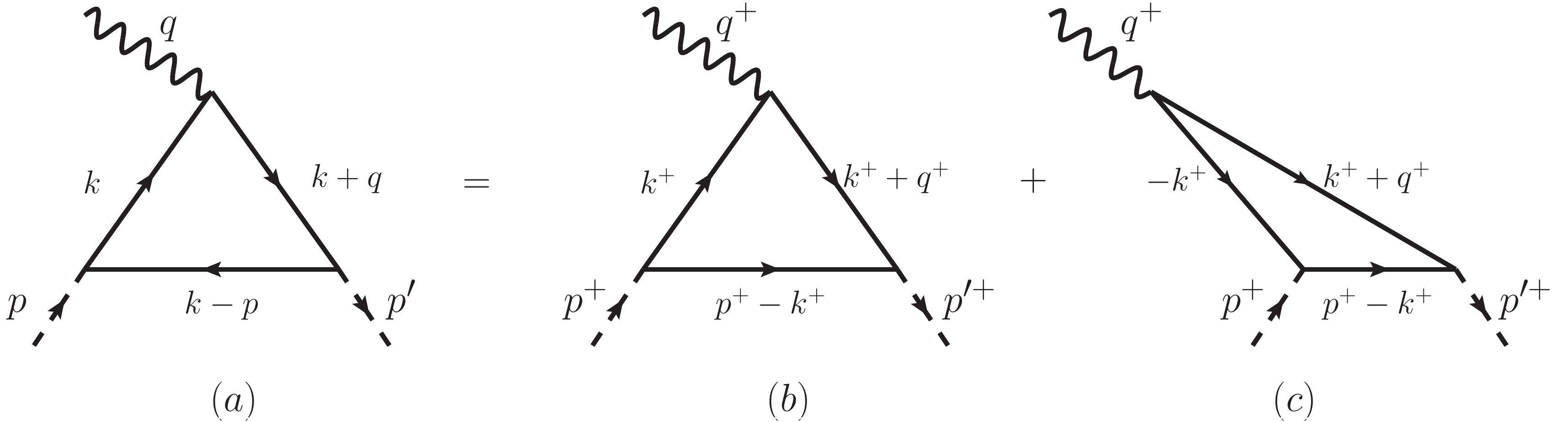}
\caption{\label{fig1}
One-loop Feynman diagrams that contribute to $\mathcal{M}(p) + \gamma^*(q) \to \mathcal{M}(p')$. 
The covariant diagram (a) is the same as the sum of the two LF time-ordered diagrams (b) and (c). 
}
\end{figure*}

The EM form factor $F_\mathcal{M}^{S}(q^2)$ of a scalar particle $\mathcal{M}$, a bosonic $q{\bar Q}$ bound state, for the process
of  $\mathcal{M}(p) + \gamma^*(q) \to \mathcal{M}(p')$ in spacelike momentum transfer ($q^2<0$) region is defined by the local 
current $J^{\mu}_{S}(0)$ through
\begin{equation}
\label{eq1}
J^{\mu}_{S}(0) = (p+p')^\mu F^{S}_{\cal M}(q^2),
\end{equation} 
where $p$ $(p')$ is the four momentum of the initial (final) state scalar particle $\mathcal{M}$  and $q=p'-p$ is 
the four-momentum transfer of the virtual photon ($Q^2 \equiv - q^2 > 0$).
The EM local current $J^{\mu}_{S}(0)$ in ($1+1$) dimensions obtained from the covariant diagram of Fig.~\ref{fig1}(a) is represented by
$J^{\mu}_{S}(0) = e_q I^\mu(m_q, m_{\bar{Q}}) + e_{\bar{Q}}I^\mu(m_{\bar{Q}}, m_q)$, where
 \begin{equation} 
 \label{eq2}
I^\mu(m_q, m_{\bar{Q}}) =i g^2 \int \frac{d^{2}k}{(2\pi)^2} \frac{2 k^{\mu}+q^{\mu}}{N_k N_{k+q} \bar{N}_{k-p}},
\end{equation}
with $N_{k} =k^{2} -m_q^{2}+i\epsilon$ and $\bar{N}_{k}=k^{2}-m_{\bar Q}^{2}+i\epsilon$ coming from 
the bosonic quark and antiquark propagators of mass $m_q$ and $m_{\bar{Q}}$, respectively, which carry the internal momentum $k$.
The normalization constant $g$ is fixed by the condition that $F^{S}_{\cal M}(q^2=0)=e_{q} +e_{\bar Q}$. 
Exchanging $m_q$ and $m_{\bar Q}$ in $I^\mu(m_q, m_{\bar{Q}})$ gives $I^\mu(m_{\bar{Q}}, m_q)$.

In LF calculations, we use the metric convention that $k \cdot q=\frac{1}{2}(k^+ q^- + k^- q^+)$.
Using this metric and choosing the plus component of the currents, $J^{+}_{S}$, the Cauchy integration over $k^-$ in Eq.~(\ref{eq2}) 
gives the two time-ordered contributions to the residue calculations, i.e., one coming from the region ${\rm S1}$ $(0 < k^+ < p^+)$  
[Fig. 1(b)] and the other from the region ${\rm S2}$ $(-q^+ < k^+ < 0)$ [Fig. 1(c)].
In the region of ${\rm S1}$ (S2), the residue is at the pole of $k^- = k^-_b$ ($k^- = k^-_r$), 
which is placed in the upper (lower) half of complex $k^-$ plane. 
Therefore, the Cauchy integration of $I^+(m_q, m_{\bar{Q}})$ in Eq.~(\ref{eq2}) over $k^-$ in ${\rm S1}$ and ${\rm S2}$ 
leads to
\begin{eqnarray}\label{eq3}
I^{+}_{\rm S1}&=& -\frac{g^2}{4\pi} \int^{p^+}_{0} dk^{+}   \frac{2 k^{+}+q^+}{C_k(k^{-}_b-k^{-}_l) (k^{-}_b-k^{-}_r)},
\nonumber\\
I^{+}_{\rm S2}&=& \frac{g^2}{4\pi} \int^{0}_{-q^+} dk^{+}   \frac{2 k^{+}+q^+}{C_k(k^{-}_{r}-k^{-}_{l})(k^{-}_{r}-k^{-}_{b})},
\end{eqnarray}
where $C_k =k^+  (k^+ + q^+) (k^+ - p^+)$ and 
\begin{eqnarray}\label{eq4}
k^{-}_{l}&=&\frac{m_q^{2}}{k^{+}}-i \frac{\epsilon}{k^{+}},\nonumber\\
k^{-}_{b}&=&p^{-}+\frac{m_{\bar Q}^{2}}{k^{+}-p^{+}}-i \frac{\epsilon}{k^{+}-p^{+}},\nonumber\\
k^{-}_{r}&=&-q^{-}+\frac{m_q^{2}}{k^{+}+q^{+}}-i \frac{\epsilon}{k^{+}+q^{+}}.
\end{eqnarray}
We note that Eq.~(\ref{eq3}) is obtained from the condition that $q^+ >0$, which means that the virtual photon is moving to the positive $z$-direction.

In the spacelike region ($q^2=q^+ q^-=-Q^2<0$) of $(1+1)$ dimensions, the momentum transfer $Q^2$ is defined as
\begin{equation}
\label{eq5}
Q^2 = M^2 {\bar\beta}^2/\beta,
\end{equation}
where $\beta = p'^+/p^+ = 1 + q^+/p^+$, ${\bar\beta}=\beta-1$, and $M^2=p^2=p'^2$.
This allows $\beta$ to have two different solutions:
\begin{equation}
\label{eq6}
\beta_\pm = \left( 1 + \frac{Q^2}{2M^2} \right) \pm \sqrt{ \left( 1 + \frac{Q^2}{2M^2}\right)^2 -1},
\end{equation}
which leads to $\beta_{\pm}=1$ (or ${\bar\beta}_{\pm}=0$)  when $Q^2=0$. 
Using the longitudinal momentum fraction $k^+= x p^+$ for the struck quark and $\beta$ for an external momentum transfer, we obtain
\begin{eqnarray}
\label{eq7}
I^+_{\rm S1} &=&  \frac{p^+ g^2}{4\pi}   \int^1_0 \frac{[dx]}{(1-x')} 
\frac{ (2 x+\bar{\beta})}{(M^2- M^2_{0})(M^2 - M'^2_{0})},
\nonumber\\
I^+_{\rm S2} &=& \frac{p^+ g^2}{4\pi}   \int^1_0 \frac{[dy]}{(1+\bar{\beta}y)} 
\frac{ (2y-1) \beta\bar{\beta}}{\left(Q^2 + \frac{m^2_q}{y(1-y)}\right) \left( M^2 - M^2_{\bar{\beta}y'}\right)},
\end{eqnarray}
where $[dZ] = \frac{dZ}{Z(1-Z)}$ and  
$ M^2_{0} = \frac{m^2_{\bar{Q}}} {1- x} + \frac{m^2_q}{x}$, $ M'^2_{0} = \frac{m^2_{\bar{Q}}} {x'} + \frac{m^2_q} {1-x'}$,
$ M^{2}_{\bar{\beta} y'} = \frac{m^2_{\bar{Q}}} {1- \bar{\beta} y'} + \frac{m^2_q} {\bar{\beta} y'}$
with  $x' = (1-x) / \beta$ and $y'=(1-y)/\beta$.
The change of variable, $x=-{\bar\beta} y$, is made to derive $I^+_{\rm S2}$ as given in Eq.~(\ref{eq7}).
While each contribution, $I^+_{\rm S1}$ and $I^+_{\rm S2}$, is independent of the choice on $\beta$, 
we take $\beta=\beta_+$ in Eq.~(\ref{eq7}) since they are obtained using the constraint $q^+>0$.
Although we do not explicitly show the results with $q^+ <0$, where the virtual photon is moving to the negative
$z$-direction, we confirmed that $I^+_{\rm S1}$ and $I^+_{\rm S2}$ are indeed independent of the
choice of $\beta$, so that $\beta=\beta_+$ and $\beta=\beta_-$ lead to the same results.

We then further evaluate the integration over the variables $(x,y)$ in Eq.~(\ref{eq7}) and combine both contributions as
$I_{\rm S}^{} = (I^+_{\rm S1}+ I^+_{\rm S2})/(p+p')^+$, to obtain the fully analytic form of $F^{S}_{\cal M}(Q^2)$ in 
$(1+1)$ dimensions as
\begin{equation}\label{eq8}
F^{S}_\mathcal{M}(Q^2) = e_q \, I^{q}_{\rm S}(Q^2) + e_{\bar{Q}} \, I^{Q}_{\rm S}(Q^2),
\end{equation}
with
\begin{eqnarray}\label{eq9}
I^{q}_{\rm S}(Q^2) &=& \frac{g^2}{8\pi  m^2_q m^2_{\bar Q}   \left(1-\omega^2 + \gamma_Q^{} \right)}
\nonumber\\
&& \mbox{} \times \left\{ C_\omega + \frac{\sqrt{1 + \gamma_Q^{}}}{\sqrt{\gamma^{}_Q}}   
\tanh^{-1}\left(\frac{\sqrt{\gamma^{}_Q}}{\sqrt{1+\gamma^{}_Q}}\right)  \right\},
\end{eqnarray}
where
\begin{eqnarray}\label{eq9-1}
C_\omega = \frac{\omega}{\sqrt{1-\omega^2} }
\left[ \tan ^{-1}(\frac{\omega_q}{\sqrt{1-\omega^2}})
+  \tan ^{-1}(\frac{\omega_Q }{\sqrt{1-\omega^2}})
\right],\nonumber\\
\end{eqnarray}
and $\omega_q =\frac{M^2 + m^2_q -m^2_Q}{2 m_q m_{\bar Q}}$, $\omega_Q =\frac{M^2-m^2_q + m^2_Q}{2 m_q m_{\bar Q}}$,
$\omega =\frac{M^2-m^2_q -m^2_Q}{2 m_q m_{\bar Q}}$ are kinematic factors.  
We note  that the momentum transfer $Q^2$ in Eq.~(\ref{eq9}) comes in only through the factor $\gamma_Q^{} = Q^2 / 4 m^2_q$.
The second part $I^Q_{\rm S} (Q^2)$ of Eq.~(\ref{eq8}) can be obtained from Eq.~(\ref{eq9}) by replacing $\gamma_Q^{}$ 
with $Q^2 / 4 m^2_{\bar Q}$.
It should be noted that our LF result in Eq.~(\ref{eq8}) is identical to the one obtained from the manifestly covariant calculations.
The analytic forms of the valence and non-valence contributions, i.e., $I^q_{\rm S1}(Q^2)$ and $I^q_{\rm S2}(Q^2)$, are  
given in Appendix. 
As one can see from Eqs.~(\ref{ap1}) and~(\ref{ap2}), the second term containing the $\tanh^{-1}$ function in Eq.~(\ref{eq9}) 
comes from the non-valence contribution.
From the form factor normalization $F^{S}_{\cal M}(Q^2)= e_q + e_{\bar Q}$, which means $I^{q}_{\rm S}(Q^2 = 0)=I^{Q}_{\rm S}(Q^2 = 0)=1$, 
we can obtain the normalization constant $g$ as
\begin{equation}\label{eq9-2}
g =\sqrt{ \frac{8\pi m^2_q m^2_{\bar Q}   (1-\omega^2) }{C_\omega + 1}}.
\end{equation}

In the limit of very high $Q^2$, the leading contribution in Eq.~(\ref{eq9}) comes from the 
$\tanh^{-1} \left(\sqrt{\frac{\gamma_Q}{1+\gamma_Q}} \right)/\gamma_Q$ term in the non-valence diagram.
Using the fact that  $\tanh^{-1} \left(\sqrt{\frac{x}{1+x}} \right)\to \ln x$ as $x\to\infty$, the leading asymptotic behavior of 
the form factor reads
\begin{equation}
\label{eq9-3}
\lim_{Q^2\to\infty} F^{S}_{\cal M}(Q^2) \sim \frac{\ln Q^2}{Q^2},
\end{equation}
which shows the power-law falloff  modified by the presence of a logarithmic function as was obtained in Ref.~\cite{GS90}.
In Ref.~\cite{Miller09}, the leading asymptotic behavior of the form factor using the same $\phi^3$ model in $(3+1)$ dimensions 
was obtained as 
$F^{S}_{\cal M}(Q^2\to\infty)\sim \ln^2 Q^2/Q^2$,
which has one more logarithmic power than the result in $(1+1)$ dimensions, and this power difference appears due to the effect of the transverse momentum.
Apart from the presence of a logarithmic function ($\sim\ln Q^2$), 
both asymptotic results in $(1+1)$-and $(3+1)$-dimensions satisfy the quark counting rules ($\sim 1/Q^2$) for the momentum transfer dependence of the
quark and antiquark bound state form factors. 
While one may avoid the nonvalence contribution using the $q^+=0$ frame in the $(3+1)$ dimensions, 
the nonvalence contribution cannot be avoided in the $(1+1)$ dimensions since $q^+\neq 0$ for $Q^2 \neq 0$.
 It turns out that the nonvalence contribution dominates in the large $Q^2$ region due to its substantial
$\ln Q^2$ behavior. This appears the characteristic of the form factor in the $(1+1)$ dimensional scalar field model that we discuss in the present work. 

\begin{figure}
\includegraphics[width=0.7\columnwidth]{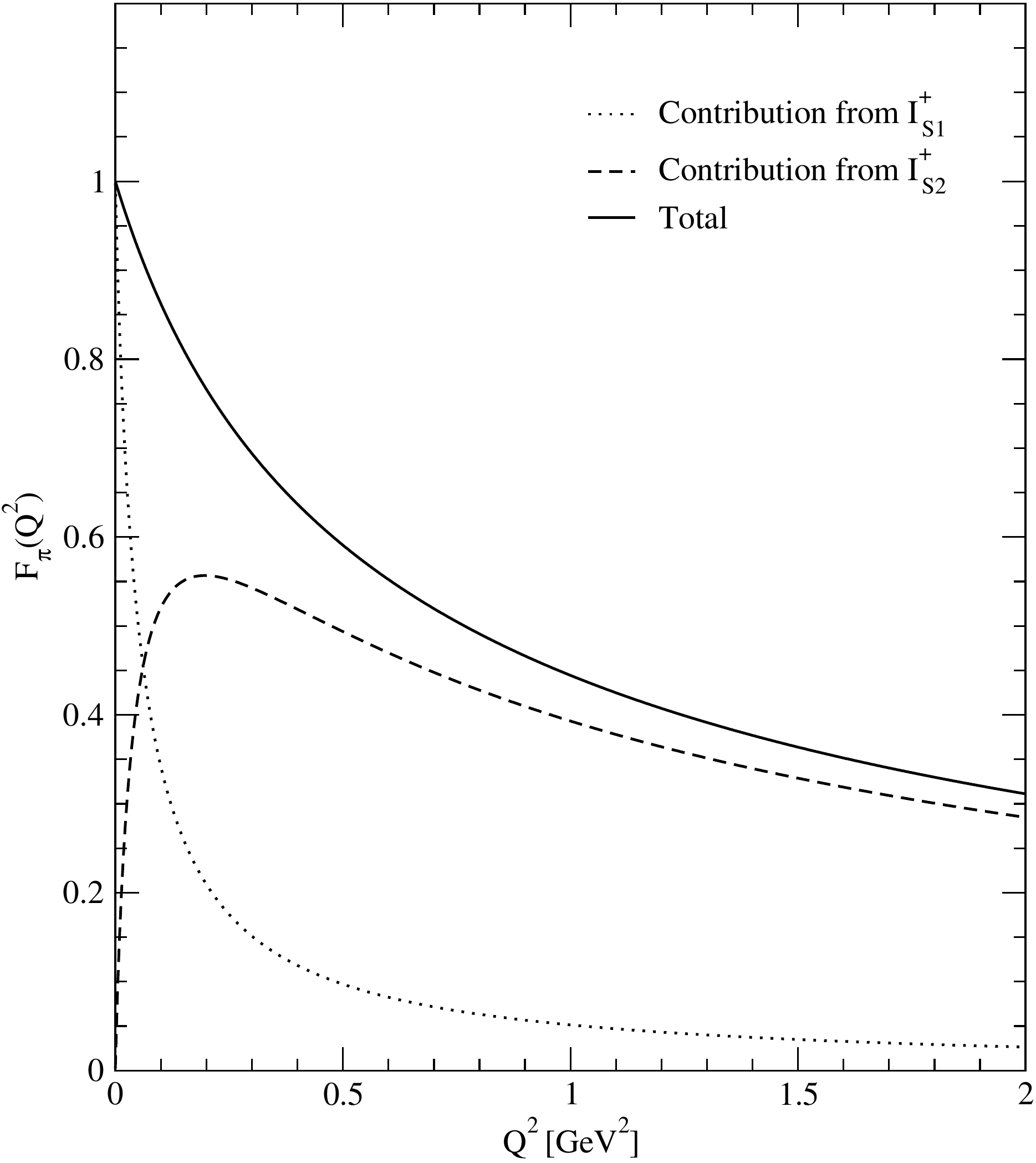}
\caption{\label{fig2} Scalar pion form factor (solid line) and its valence  (dotted line)
 and non-valence  (dashed line) contributions for $0\leq Q^2\leq 2$ GeV$^2$ region.}
\end{figure}

Shown in Fig.~\ref{fig2} is the EM form factor for a `scalar pion' in the spacelike region of $0\leq Q^2\leq 2$~GeV$^2$. 
In this model calculation, we use $M=0.14$ GeV and $m_q=m_{\bar Q}=0.25$~GeV.  
The dotted, dashed, and solid lines represent the valence contribution $I^+_{\rm S1}$,  non-valence contribution $I^{+}_{\rm S2}$, 
and the total result of the form factor, respectively. 
We find that, while the valence contribution dominates for small $Q^2$ region, the non-valence contribution takes over the valence one
for $Q^2\geq 0.1$~GeV$^2$ and most of the contribution to the form factor for high $Q^2$ comes from the non-valence diagram,
indicating significant contributions from the higher-Fock components.

\subsection{Form factor in the timelike region}

\begin{figure*}
\includegraphics[height=3.5cm, width=13cm]{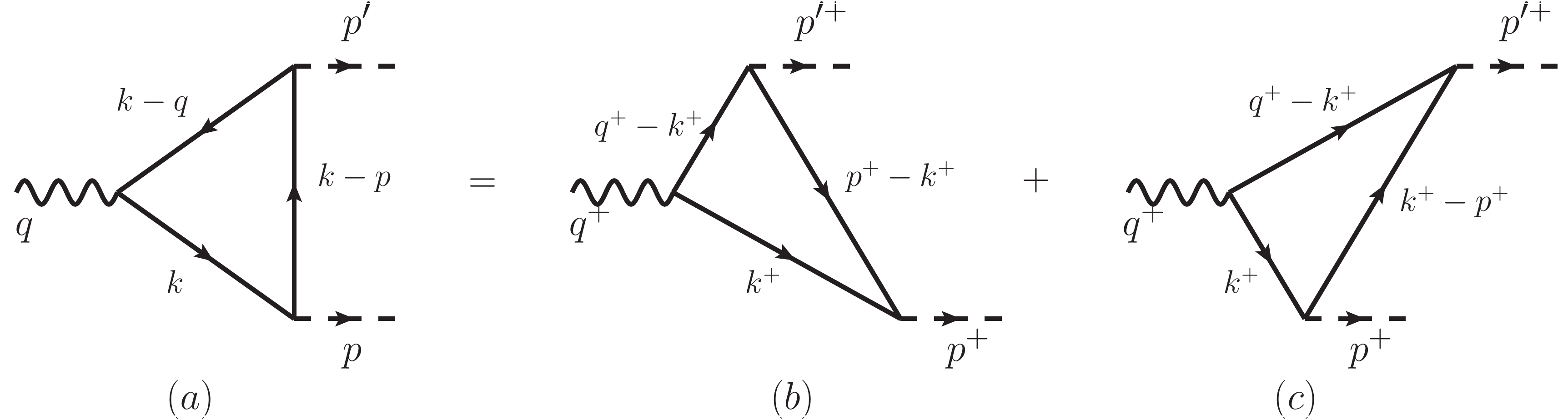}
\caption{\label{fig3}
One-loop Feynman diagrams that contribute to $\gamma^*(q) \to \bar{\mathcal{M}}(p)+ {\cal M}(p')$. The covariant diagram (a) is 
the same as the sum of the two LF time-ordered diagrams (b) and (c). 
}
\end{figure*}

The process of  $\mathcal{M}(p) + \gamma^*(q) \to  \mathcal{M}(p')$ in spacelike momentum transfer ($q^2<0$) region 
can be made to the timelike $(q^2>0)$ process such as $\gamma^*(q)\to \bar{\mathcal{M}}(p) + \mathcal{M}(p')$, i.e.,
the reaction for a virtual photon decaying into a $q{\bar Q}$ bound state scalar particle $\mathcal{M}$ and its antiparticle 
$\bar{\mathcal{M}}$ from the principle of crossing symmetry.
Therefore, the timelike EM form factor $F^{T}_{\cal M}(q^{2})$ for $\gamma^*(q) \to \bar{\mathcal{M}}(p) + {\mathcal{M}}(p')$
is defined by the local current $J^{\mu}_T(0)$ as
\be\label{eq10}
J^{\mu}_T(0) =(p-p')^{\mu}~F^{T}_{\cal M}(q^{2}), 
\ee
where $q = p + p'$ is the four momentum transfer of the virtual photon satisfying $q^2 = q^+ q^->0$.
The covariant diagram describing $\gamma^*(q)\to \bar{\cal M}(p) + {\cal M}(p')$ process is shown in Fig.~\ref{fig3}(a) 
and the local current is obtained by $J^{\mu}_{T}(0) = e_q I^\mu(m_q, m_{\bar{Q}}) + e_{\bar{Q}}I^\mu(m_{\bar{Q}}, m_q)$
as in the case of $\mathcal{M}(p) + \gamma^*(q) \to \mathcal{M}(p')$ process. 
Essentially, $I^\mu(m_q, m_{\bar{Q}})$ in this timelike process has the same form given by Eq.~(\ref{eq2})
but with the overall sign changed.

In LF calculations, using the plus component of the currents $J^{+}_{T}$, the Cauchy integration over $k^-$ in 
Eq.~(\ref{eq2}) gives two time-ordered
contributions  to the residue calculations:
one coming from the region ${\rm T1}$ $(0< k^+ < p^+)$ [Fig. ~\ref{fig3}(b)] and
the other from the region ${\rm T2}$ $(p^+ < k^+ < q^+)$ [Fig.~\ref{fig3}(c)].
In the region of  T1 (T2), the residue is at the pole of $k^- = k^-_l$ ($k^- = k'^-_r$), 
where $k'^{-}_r$ is the same as $k^{-}_r$ but with replacing $q$ by $-q$. 
The poles $k_l^-$ and $k'^{-}_r$ are in the lower and upper half of the complex-$k^-$ plane, respectively.
This allows one to obtain the Cauchy integration of $I^+(m_q, m_{\bar{Q}})$ in Eq.~(\ref{eq2}) over $k^-$ 
in the regions T1 and T2  as
\begin{eqnarray}
\label{eq11}
I^{+}_{\rm T1}&=& \frac{g^2}{4\pi} \int^{p^+}_{0} dk^{+}   \frac{2 k^+ - q^+}{C'_k(k^{-}_{l}-k^{-}_{b})(k^{-}_{l}-k'^{-}_{r})},
\nonumber\\
I^{+}_{\rm T2}&=& -\frac{g^2}{4\pi} \int^{q^+}_{p^+} dk^{+}   \frac{2 k^+ - q^+}{C'_k(k'^{-}_r-k^{-}_l) (k'^{-}_r-k^{-}_b)},
\end{eqnarray}
where $C'_k=C_k (q \rightarrow -q)$.

In the timelike region of $(1+1)$ dimensions, the momentum transfer $q^2$ is defined by
\begin{equation}
\label{eq12}
q^2 = M^2 (1+\al)^2/\al,
\end{equation}
where $\al = p'^+/p^+=q^+/p^+ -1$ and the two solutions for $\al$ are given by
\begin{equation}
\label{eq13}
\al_\pm = \left( \frac{q^2}{2M^2} -1\right) \pm \sqrt{ \left( \frac{q^2}{2M^2} -1\right)^2 -1}.
\end{equation} 
This shows that both $\alpha_{\pm}=1$ correspond to the threshold $q^2 = 4 M^2$. 
However, the EM form factor $F^{T}_\mathcal{M}(q^{2})$ is independent of the subscript sign of $\alpha$ as in the case of
$F^{S}_\mathcal{M}(Q^2)$ given by Eqs.~(\ref{eq8}) and~(\ref{eq9}), which can be seen below.

With $\alpha$ and $k^+ = x p^+$, we rewrite Eq.~(\ref{eq11}) as
\begin{eqnarray} \label{eq14}
I^+_{\rm T1} &=&  -\frac{p^+ g^2}{4\pi}  \int^1_0  \frac{[dx]}{1 + \al - x} 
\frac{ \al (1+\al - 2x)/(1+\al)}{(M^2-M^2_{0x}) \left[ M^2 - \frac{\al m^2_q}{x (1+\al -x)} \right]},
\nonumber\\
I^+_{\rm T2} &=& \frac{p^+ g^2}{4\pi}   \int^1_0 \frac{[dx']}{1 + \al x'} 
\frac{\al (2\al x' - \al +1 )/(1+\al)}{(M^2-M'^2_{0x}) \left[ M^2 - \frac{m^2_q}{(1-x') (1+\al x')} \right]},
\nonumber\\
\end{eqnarray} 
where $M^2_{0x} = \frac{m^2_q}{x}  + \frac{m^2_{\bar Q}}{1-x}$ and $M'^2_{0x}=M^2_{0x} (m_q \leftrightarrow m_Q)$. 
The change of variable, $x \equiv \al x'+1$, is made in the calculation of $I^+_{\rm T2}$.

Then we further integrate over the variables $(x, x')$ to combine both contributions, 
$I_{\rm T} = (I^+_{\rm T1}+ I^+_{\rm T2})/(p-p')^+$,
which leads to the fully analytic form of $F^{T}_{\cal M}(q^2)$ in $(1+1)$ dimensions as
\begin{equation}
\label{eq15}
F^{T}_{\cal M}(q^2) = e_q I^q_{\rm T} (q^2) + e_{\bar Q} I^{Q}_{\rm T} (q^2),
\end{equation}
where
\begin{eqnarray}
\label{eq16}
I^q_{\rm T}(q^2) &=& \frac{ g^2}{8\pi  m^2_q m^2_{\bar Q} \left(1 - \omega^2 - \gamma_q \right)}
\nonumber\\
&& \mbox{} \times \left\{
C_\omega +
\sqrt{\frac{1-\gamma_q}{\gamma_q}}  
\tan^{-1}\left( \frac{1}{\sqrt{\frac{1-\gamma_q}{\gamma_q}}} \right)
 \right\},
\end{eqnarray} 
with $\gamma_q = q^2 / 4m^2_q$. 
The second part $I^Q_{\rm T} (q^2)$ of Eq.~(\ref{eq15}) can be obtained from Eq.~(\ref{eq16}) by replacing $\gamma_q$ 
with $q^2 / 4m^2_{\bar Q}$.
It can be easily seen that Eq.~(\ref{eq16}) and Eq.~(\ref{eq9}) are essentially the same through the analytic continuation 
from spacelike $Q^2$ region to timelike $q^2(=-Q^2)$ region.
This shows that the timelike EM form factor can be analytically continued to spacelike region by changing $q^2\to -q^2$ 
in the form factor and vice versa.

It should be noted that the threshold points of the timelike form factor depend on the bound state condition.
That is, the denominator factor $(1 - \omega^2 - \gamma_q)$ in Eq.~(\ref{eq16}) is always nonzero 
for strong bound state satisfying both  $M< m_q + m_{\bar Q}$ and $M^2 < m^2_q + m^2_{\bar Q}$.
In this strong bound state case,  the imaginary part of $I^{q(Q)}_{T}(q^2)$ starts to develop at  $q^2 \geq 4 m^2_{q(Q)}$. 
In other words, the thresholds for strong bound state are given by the ``normal" threshold points, i.e.,
$q^2_{\rm min}= 4m^2_q$ and $4m^2_{\bar Q}$ for $\gamma^* q{\bar q}$ and $\gamma^* Q{\bar Q}$ vertices, respectively.
Furthermore, one can easily extract the analytic form for the imaginary part of the timelike form factor from Eq.~(\ref{eq16})
as
\begin{equation}
\label{eq17}
\mbox{Im}[I^q_{\rm T}(q^2)] 
=-\frac{ g^2  \theta(\gamma_q -1)}{16 m^2_q m^2_{\bar Q} \left(1 - \omega^2 - \gamma_q \right)}\sqrt{\frac{\gamma_q-1}{\gamma_q}},
\end{equation}
where $\theta(\gamma_q-1)$ is the Heaviside step function, i.e., $\theta (\gamma_q-1)=1$ for $\gamma_q >1$ and vanishes otherwise.

On the other hand, for the weakly bound state satisfying $M< m_q + m_{\bar Q}$ but $M^2 > m^2_q + m^2_{\bar Q}$,
the singular $q^2$ point, satisfying $1 - \omega^2 - \gamma_q=0$, exists. 
Because of this, the singularity for the weakly bound state starts at the ``anomalous" threshold points,
\begin{equation} \label{eq19}
q^2_{\rm min} = \frac{1}{m^2_{{\bar Q}(q)}} [m^2_{q ({\bar Q})} - (M - m_{{\bar Q}(q)})^2]
[ (M + m_{{\bar Q}(q)})^2 - m^2_{q ({\bar Q})}],
\end{equation} 
for $\gamma^* q{\bar q}\; (\gamma^* Q{\bar Q})$ vertex prior to the normal thresholds.
Our result for the anomalous thresholds given by Eq.~(\ref{eq19}) is exactly the same as the one in Ref.~\cite{Gasiorowicz} for the analysis 
of the one-particle matrix elements of  a scalar current.
Especially, in this weakly bound state in $(1+1)$ dimensions, the real part of the timelike form factor diverges at the anomalous threshold.

As a consistency check, we compare our direct result for 
the form factor with the dispersion relations (DRs) given by 
\begin{eqnarray}\label{eq18}
{\rm Re}[F(q^{2})]=\frac{1}{\pi} \mathcal{P} \int_{-\infty}^{\infty} dq'^{2} ~\frac{{\rm Im}[F(q'^{2})]}{q'^{2}-q^{2}},
\nonumber\\
{\rm Im}[F(q^{2})]=-\frac{1}{\pi} \mathcal{P} \int_{-\infty}^{\infty} dq'^{2} ~\frac{{\rm Re}[F(q'^{2})]}{q'^{2}-q^{2}},
\end{eqnarray}
where ${\mathcal P}$ indicates the Cauchy principal value. 
For the strong bound state case, we confirm that the real part of the form factor obtained from 
the direct calculation in Eq.~(\ref{eq16}) is exactly the same as the one obtained from DR  using 
the analytic form of the imaginary part given by Eq.~(\ref{eq17}).
However, for the weakly bound state case, we note the importance of taking into account 
the infinitesimal width $\Gamma$ as $(1-\omega^2 - \gamma_q)\to (1-\omega^2 - \gamma_q- i\Gamma)$ in Eq.~(\ref{eq16}) 
for the timelike form factor in order to remedy the singularity at the anomalous threshold point.
With this care, we explicitly obtain both real and imaginary part of the timelike form factor for the weakly bound state as
\begin{eqnarray}
{\rm Re}[I^q_{\rm T}] &=& \frac{ g^2}{8 \pi  m^2_q m^2_{\bar Q} \left((1 - \omega^2 - \gamma_q)^2 + \Gamma^2\right)}\nonumber\\
&&  \times\left\{ 
(1-\omega^2 - \gamma_q)  {\tilde C}^q_\omega
-\frac{\pi}{2} \Gamma\sqrt{\frac{\gamma_q -1}{\gamma_q}} ~\theta(\gamma_q-1) \right\},
\nonumber \end{eqnarray} \begin{eqnarray}
{\rm Im}[I^q_{\rm T}] &=& \frac{ g^2}{8 \pi  m^2_q m^2_{\bar Q} \left((1 - \omega^2 - \gamma_q)^2 + \Gamma^2\right)}\nonumber\\
&& \times \left\{
 \Gamma  {\tilde C}^q_\omega
-(1-\omega^2 - \gamma_q) ~\frac{\pi}{2} \sqrt{\frac{\gamma_q -1}{\gamma_q}} \theta(\gamma_q-1) 
 \right\},
\nonumber\\
\label{eq19-1}
\end{eqnarray} 
where
\begin{equation} \label{eq19-2}
{\tilde C}^q_\omega = C_\omega +
{\rm Re}\left[\sqrt{\frac{\gamma_q -1}{\gamma_q}}  
\tanh^{-1}\Bigg( \frac{1}{\sqrt{\frac{\gamma_q -1}{\gamma_q}}} \Bigg)\right].
\end{equation}

\begin{figure}
\includegraphics[width=0.7\columnwidth]{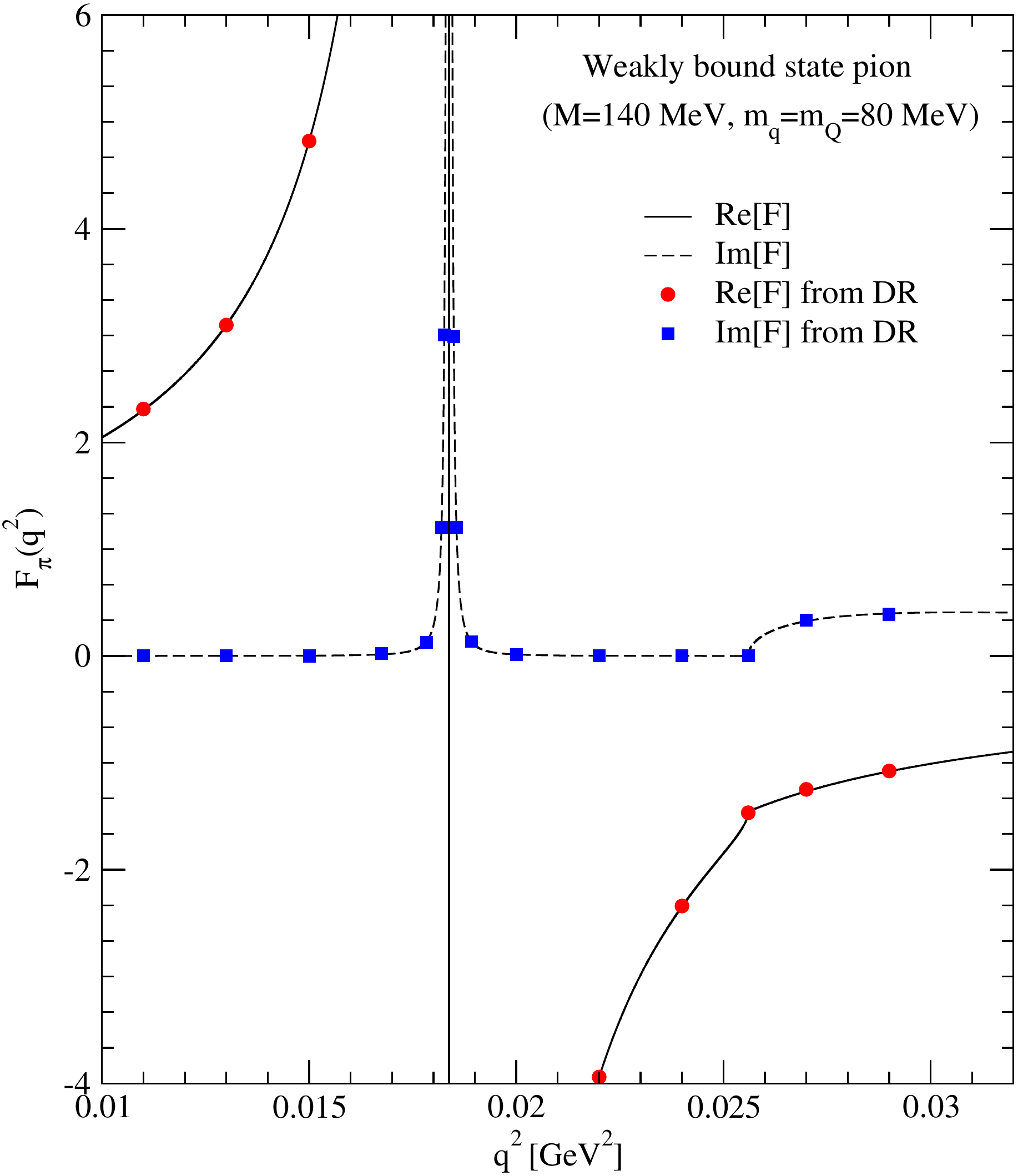}
\caption{\label{fig4} Timelike form factor for the weakly bound state pion obtained with $M=0.14$ GeV and $m_q=m_{\bar Q}=0.08$ GeV. 
The direct results of real (solid line) and imaginary (dashed line) parts of the form factor are compared with the real (circle)
and imaginary (square) parts obtained from the dispersion relations.}
\end{figure}

For an explicit demonstration, we show in Fig.~\ref{fig4} the timelike form factor for the weakly bound state
pion obtained with $M=0.14$~GeV and $m_q=m_{\bar Q}=0.08$~GeV together with an infinitesimal width $\Gamma =0.0001$~GeV.
In this case, the normal threshold ($q^2=4 m^2_q$) appears
at $q^2 = 0.0256$ GeV while the anomalous threshold defined in Eq.~(\ref{eq19}) appears at $q^2 \simeq 0.0184$~GeV.
The solid and dashed lines are the results for the real and imaginary parts of the form factor obtained from Eq.~(\ref{eq19-1}).
The circle and square data represent the results for the real and imaginary parts of the form factor obtained from the DRs given by Eq.~(\ref{eq18}).
This confirms that our direct results presented in Eq.~(\ref{eq19-1}) are in complete agreement with the DRs.

\section{Longitudinal charge density and charge radius}
\label{sec:3}

The intrinsic size of the hadron is defined through the slope of the form factor, i.e.,
the mean-square charge radius in three dimensional space is given by
$\braket{r^2} = -6 dF(Q^2)/dQ^2|_{Q^2=0}\equiv -6 F'(0)$.
The size of the hadron may also be computed from the intrinsic charge density $\rho ({\bf r})$ 
in three dimensional space within a nonrelativistic theory, defined by the Fourier transform of the form factor,
\begin{equation}
\label{eq23c}
 \rho ({\bf r}) = \frac{1}{(2\pi)^3} \int d^3Q\; e^{-i{\bf Q}\cdot{\bf r}} F(Q^2).
\end{equation}
At relativistic energies this interpretation becomes obscured because of its dependence on the reference 
frame~\cite{Hofstadter56,GR,Burkardt00,Miller09,Miller09a}. 
With this caveat,  the mean-square charge radius $\braket{r^2}$ may be defined as the second moment 
of the intrinsic charge density $\rho ({\bf r})$, which gives  $\braket{r^2} = \int r^2 \rho ({\bf r}) d^3r$.

However, since the intrinsic charge density is inherently nonrelativistic and requires relativistic corrections,
the transverse charge density $\rho(b)$ was proposed in Refs.~\cite{Burkardt02, Diehl02,Miller09,Miller09a}
as the true charge density without the need for relativistic corrections, which is obtained by the following 
two-dimensional Fourier transform,
\begin{equation}
\label{eq24c}
 \rho (b) = \frac{1}{(2\pi)^2} \int d^2{\bf q}_\perp\; e^{-i{\bf q_\perp}\cdot{\bf b}} F(Q^2={\bf q}^2_\perp),
\end{equation}
where ${\bf b}$ is the two-dimensional transverse variable and $F(Q^2={\bf q}^2_\perp)$ is obtained from the
DYW frame (i.e., $q^+=0$ and $q^2 =-{\bf q}^2_\perp=-Q^2$). 
This transverse density is also the integral of the three dimensional infinite-momentum frame (IMF) density $\rho(x^-, b)$
over all values of the longitudinal position coordinate~\cite{Miller09a}.
The central charge density of the hadron is determined by $\rho(b=0)$ because of the Lorentz-contraction of the 
longitudinal dimension in the IMF.
The mean-square transverse radius $\braket{b^2}$ is then given in terms of $\rho(b)$ as
$\braket{b^2} = \int d^2 b\, b^2\rho(b) = - 4F'(0)$.

While the frame dependence of the intrinsic charge density in the longitudinal direction due to relativistic corrections, 
such as the Lorentz contraction, has been widely discussed in Refs.~\cite{Burkardt00,Miller09,Miller09a}, the explicit 
estimation of the relativistic effect is yet to be fully discussed.
The purpose of this section is to apply our results for the EM form factor in $(1+1)$ dimensions to obtain the 
longitudinal charge density $\rho(r_z)$ and to understand the difference between the intrinsic charge density 
obtained from the one-dimensional Fourier transform of the charge form factor $F(Q^2)$
and the relativistic version of the true static charge density obtained from the $(1+1)$-dimensional Fourier transform
of $J^0(Q^2)=(p+p')^0 F(Q^2)$ in the so-called ``Breit frame (BF)" where no energy is transferred to the hadron, 
i.e., $q=(q^0, q_z)=(0, Q)$, $p=(E, p_z)$, and $p'=(E, -p_z)$.
In LFD, the longitudinal charge density is discussed in terms of the boost invariant variable $\tilde z = p^+ x^-$.

\subsection{Intrinsic longitudinal charge density}
The intrinsic longitudinal charge density (ILD)  $\rho_{\rm ILD}^{} (r_z)$ may be defined 
by the one-dimensional Fourier transform of the spacelike form factor $F^{S}_\mathcal{M}(Q^2) \equiv F(Q^2)$ as
\begin{equation}
\label{eq25c}
 \rho_{\rm ILD}^{} (r_z)= \frac{1}{2\pi} \int dQ\; e^{-iQ r_z} F(Q^2),
\end{equation}
where $Q$ corresponds to the longitudinal component (i.e., $Q=q_z$) of momentum transfer.
The intrinsic longitudinal density $\rho_{\rm ILD}^{} (r_z)$ represents the probability that electric charge is
located at a longitudinal distance $r_z$ from the longitudinal center of momentum with the normalization 
condition $\int dr_z  \, \rho_{\rm ILD}^{} (r_z)=F(0)$.

Using the inverse Fourier transform and the normalization of $\rho_{\rm ILD}^{}(r_z)$, we obtain
\begin{eqnarray}
\label{eq26c}
F(Q^2) &=&  \int dr_z\; e^{iQ r_z}  \rho_{\rm ILD}(r_z)
\nonumber\\
&=& 1- \frac{1}{2} Q^2 \braket{r^2_z}_{\rm ILD} + \cdots,
\end{eqnarray}
where  $\braket{r^2_z}_{\rm ILD} = \int dr_z\; r^2_z  \rho_{\rm ILD}^{} (r_z)$.
From Eqs.~(\ref{eq8}) and~(\ref{eq9}), we can explicitly obtain the analytic form of the mean square charge radius  
in the longitudinal direction $r_z$, i.e., $\braket{r^2_z}_{\rm ILD} = -2~ \partial F/ \partial Q^{2} \vert_{Q^2=0}$, as
\begin{equation}
\label{eq27c}
\braket{r^2_z}_{\rm ILD} =  \frac{1}{6} \left[ \frac{3}{1-\omega^2} - \frac{1}{1+C_\omega} \right] 
\left( \frac{e_q}{m^2_q} +\frac{e_{\bar Q}}{m^2_{\bar Q}} \right).
\end{equation}
It is interesting to compare our result with that obtained from the simple analysis of nonrelativistic quark model~\cite{Green}, 
where the mean-square charge radius, $\braket{r^2_{\rm em}} = -6 F'(0)$, defined in three spatial dimensions is obtained as
$\braket{r^2_{\rm em}} = \frac{(e_q m^2_{\bar Q} + e_{\bar Q} m^2_q) \braket{\delta^2} }{(m_q + m_{\bar Q})^2}$ with
$\bm{\delta}={\bf r}_q - {\bf r}_{\bar Q}$ being relative coordinate. 
This result was derived from $\braket{r^2_{\rm em}} = \braket{ \sum^{2}_{i} e^2_i (r_i -R)^2 }$,  i.e., the deviation from 
the center-of-mass position $R$ squared weighted by the charge of the quark and antiquark constituents. 
While $\braket{ r^2 }_{\rm  ILD}$ and $\braket{r^2_{\rm em}}$ were derived from different spacetime dimensions and different methods,
the both have the common factor $e_q m^2_{\bar Q} + e_{\bar Q} m^2_q$. 
From this common factor in charge radius, it is easy to find that the neutral meson such as  $K^{0}(d{\bar s})$
has a negative square charge radius.%
\footnote{A negative value for $\braket{r^2_{\rm em}}$ happens when the lighter negatively charged $d$ quark is orbiting 
around the heavier ${\bar s}$ quark~\cite{Green}. 
}

Of particular interest, we also obtain $\braket{r^2_z}_{\rm ILD}$ in terms of binding energy $B$ defined by $M= 2m_q -B$
for equal constituent mass case ($m_q = m_Q$), which leads to
\begin{widetext}
\begin{equation}
\label{eq27d}
\braket{r^2_z}_{\rm ILD} = \frac{1}{6 m^2_q}
\biggl[
\frac{12}{\varepsilon (4-\varepsilon)(2-\varepsilon)^2}
-\frac{\sqrt{\varepsilon(4-\varepsilon)}(2-\varepsilon)}{ \sqrt{\varepsilon(4-\varepsilon)}(2-\varepsilon) 
+ 2(\varepsilon^2 - 4\varepsilon +2) \tan^{-1} \left[\frac{2-\varepsilon}{\sqrt{\varepsilon(4-\varepsilon)}}\right]}
\biggr],
\end{equation}
\end{widetext}
where $\varepsilon = B / m_q$  is the dimensionless parameter ranging from zero binding ($B =0$) to 
maximal binding (i.e., $B=2 m_q$) limits. While $\braket{r^2_z}_{\rm ILD}\to\infty$ in the zero binding limit, 
it decreases monotonically to the minimum value $\braket{r^2_z}_{\rm ILD}\to 1/ (5 m^2_q)$ as $B\to 2 m_q$, 
which is consistent with the observation made in Ref.~\cite{GS90}.
As one can see, the charge radius is getting smaller as the constituent mass $m_q$ is getting larger.

\subsection{Relativistic longitudinal charge density in BF}

In $(1+1)$ dimensions, the Fourier transform of the current $J^\mu$ is given by
\begin{equation}
\label{app1}
{\tilde J}^\mu(t, r_z) =\frac{1}{(2\pi)^2}\int d^2q\; J^\mu(q^2) e^{i (q^0 t - q^3 r_z)}.
\end{equation} 
If we take the BF, where $q^{0}=0$ and $q^{2}=-(q^3)^{2}=-Q^2$, the momentum of incoming meson, $p$, and 
that of the outgoing meson, $p'=p + q$, are given by
\begin{eqnarray}
\label{app2}
(p^0, p^3)_{\rm BF} &=& \frac{1}{2} \left( \sqrt{4 M^2 +Q^2}, -\sqrt{Q^2}\right),
\nonumber\\
(p'^{0}, p'^3)_{\rm BF} &=&  \frac{1}{2}  \left( \sqrt{4 M^{2} + Q^2}, \sqrt{Q^2} \right).
\end{eqnarray}
In the BF, only the time component of the currents $J^\mu$ in Eq.~(\ref{eq1}) survives and the space component is zero
so that
\begin{eqnarray}
\label{app3}
J^{0}_{\rm BF}(q^{2})&=&\sqrt{4 M^{2} + Q^{2}}\;F(q^{2}),
\nonumber\\
J^{3}_{\rm BF}(q^{2})&=&0.
\end{eqnarray}
Then the Fourier transform of the current $J^0$ in the BF results in
\begin{equation}
\label{app4}
\rho(r_z) =\frac{1}{2\pi}\int dQ\; {\cal F}(Q^2) e^{-i Q r_z} ,
\end{equation}
where $\rho(r_z) =\int dt\, {\tilde J}^0 (t, r_z)$ corresponds to the longitudinal charge density  and
${\cal F}(Q^2) \equiv \sqrt{4 M^{2} + Q^{2}}\;F(Q^2)$.  
It should be noted that the prefactor $\sqrt{4 M^{2} + Q^{2}}$ in ${\cal F}(Q^2)$ depends on the reference frame 
while the form factor $F(Q^2)$ is Lorentz-invariant.
Since Eq.~(\ref{app4}) leads to the normalization of $\rho(r_z)$ as $\int dr_z\;\rho (r_z) = {\cal F}(0) = 2M$, we 
redefine Eq.~(\ref{app4}) as
\be\label{app5}
\rho_{\rm BF}^{} (r_z) =\frac{1}{2\pi}\int dQ\; {\cal F}_{\rm BF}^{} (Q^2) e^{-i Q r_z} ,
\ee
where $\rho_{\rm BF}^{} (z) =\rho(z) / 2M$ and ${\cal F}_{\rm BF}^{} (Q^2) = {\cal F}(Q^2) / 2M$. 
Equation~(\ref{app5}) now satisfies $\int dr_z\;\rho_{\rm BF}^{} (r_z) =1$ and ${\cal F}_{\rm BF}^{} (0)=1$.

Using the inverse Fourier transform and the normalization of $\rho_{\rm BF}^{}(r_z)$, we obtain
\begin{eqnarray}
\label{app6}
{\cal F}_{\rm BF}^{} (Q^2) &=&  \int dr_z\; e^{iQ r_z}  \rho_{\rm BF}^{} (r_z)
\nonumber\\
&=& 1- \frac{1}{2} Q^2 \braket{r^2_z}_{\rm BF} + \cdots,
\end{eqnarray}
where $\braket{ r^2_z }_{\rm BF} = \int dr_z\; r^2_z  \rho_{\rm BF}^{} (r_z)$.
Comparing to $\braket{ r^2_z }_{\rm ILD} $, we find
\be\label{app8}
\braket{r^2_z }_{\rm BF}  = \braket{ r^2_z }_{\rm ILD} - \frac{1}{4M^2}.
\ee
We note that $\braket{r^2_z }_{\rm BF}$ is independent of time $t$ and is smaller than $\braket{r^2_z }_{\rm ILD}$ 
due to the Lorentz contraction. 
For instance, if the mass of the bound state is  $M \simeq 1~\mbox{GeV}$,  then 
$\braket{r^2_z }_{\rm BF}  = \braket{r^2_z}_{\rm ILD} - 0.01~\mbox{fm}^2$ and the relativistic correction is getting
larger (smaller) as $M$ gets smaller (larger) as expected.

\begin{figure}[t]
\includegraphics[width=0.7\columnwidth]{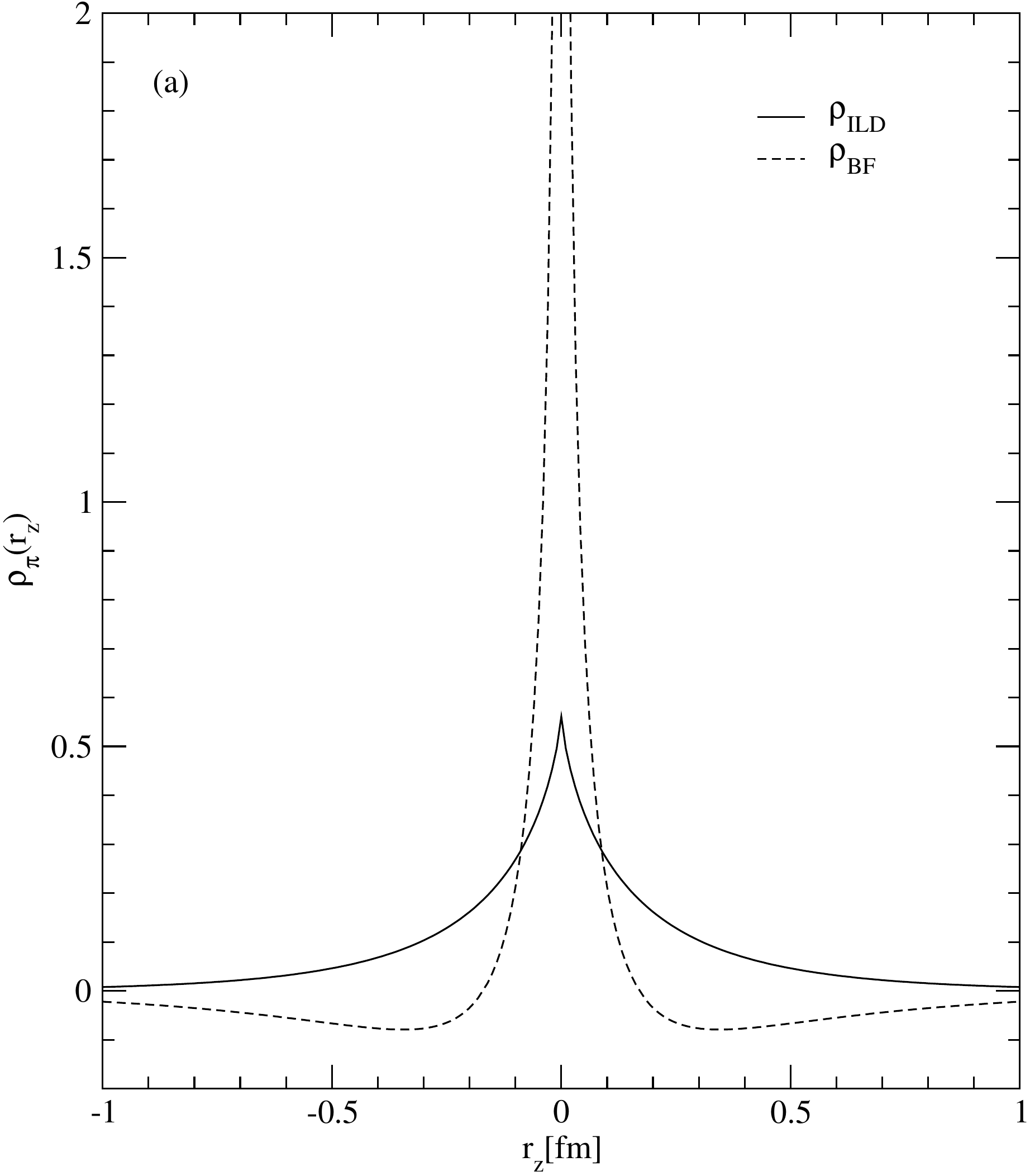}
\includegraphics[width=0.7\columnwidth]{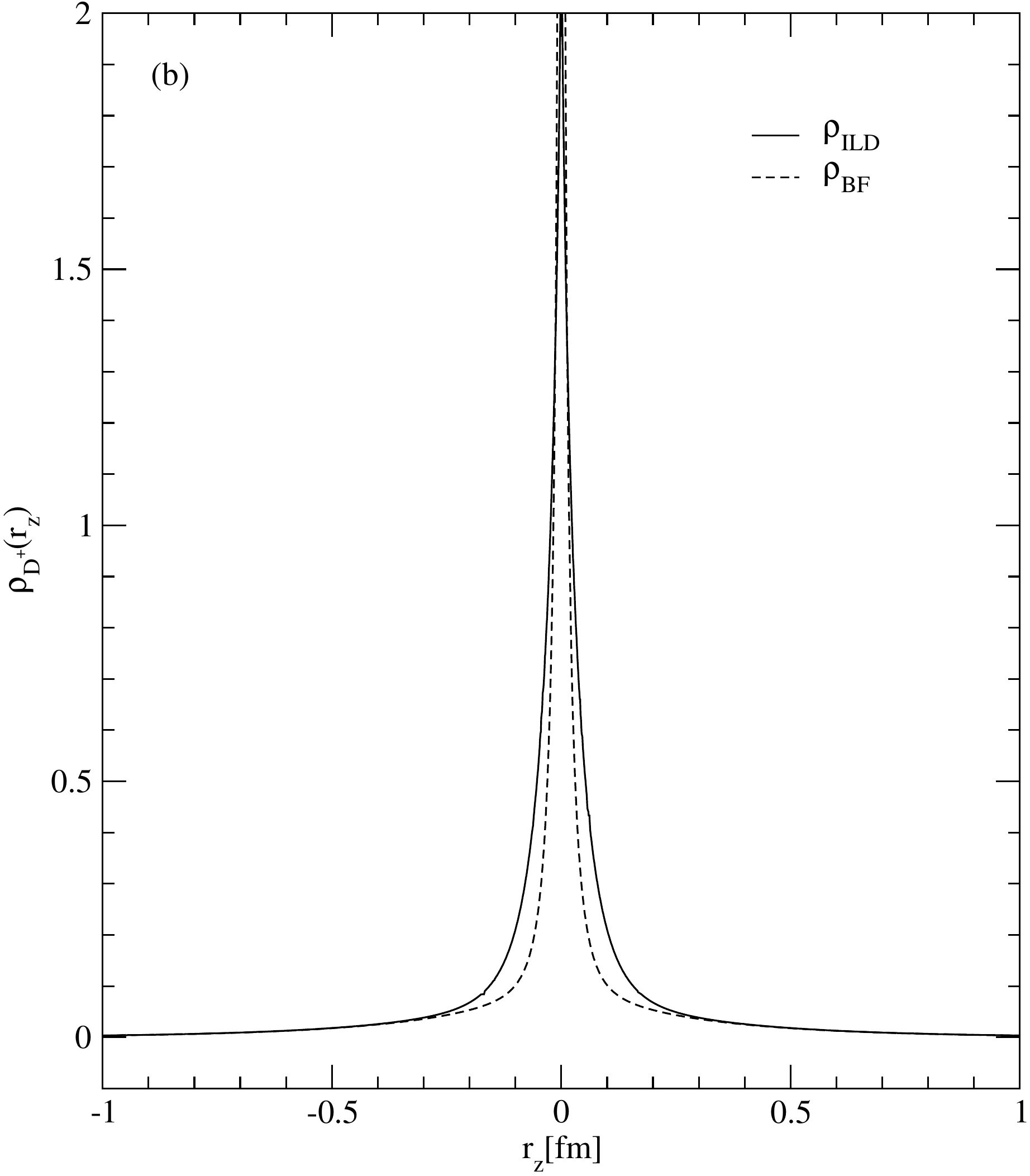}
\caption{\label{fig5} Comparison of $\rho_{\rm ILD}^{}(r_z)$ and $\rho_{BF}^{}(r_z)$ for  the strongly bound state $\pi^+$ in (a)  
and the weakly bound state $D^+$ meson in (b). }
\end{figure}

In Fig.~\ref{fig5}, we compare the two longitudinal charge densities, $\rho_{\rm ILD}^{}(r_z)$ (solid line) and 
$\rho_{\rm BF}^{}(r_z)$ (dashed line), for a strongly bound state $\pi^+$ [Fig.~\ref{fig5}(a)] and a weakly bound 
state $D^+$ meson [Fig.~\ref{fig5}(b)] in the range of $-1\leq r_z \leq 1$ fm.  
These numerical results are estimated with $m_u=m_d = 0.25~\mbox{GeV}$ and $m_c = 1.8~\mbox{GeV}$ together 
with the physical meson masses, i.e., $M_{\pi^+}= 0.14~\mbox{GeV}$ and $M_{D^+}= 1.870~ \mbox{GeV}$.
One can clearly see that  $\rho_{\rm BF}^{}(r_z)$ are more narrowly peaked near the longitudinal center of the momentum 
than $\rho_{\rm ILD}^{}(r_z)$ due to the Lorentz contraction, which is more significant for $\pi^+$ than for $D^+$.

Other reference frame may be obtained by the Lorentz transformation from the BF. 
For instance, the target rest frame (TRF), where $(p^0, p^3)_{\rm TRF}=(M,0)$, can be obtained from the BF
by the following Lorentz transformation:
\begin{equation}
\label{app8}
\left( \begin{array}{cc} 
\gamma & \gamma\beta  \\
\gamma\beta & \gamma  \\
\end{array} \right) 
\left( \begin{array}{c} 
p^{(\prime) 0}_{\rm BF} \\
p^{(\prime) 3}_{\rm BF} \\
\end{array} \right) =
\left( \begin{array}{c} 
p_{\rm TRF}^{(\prime)0}\\
p_{\rm TRF}^{(\prime)3} \\
\end{array} \right),
\end{equation}
where the Lorentz factors are given by $\gamma = \frac{\sqrt{4 M^{2} + Q^2}}{2 M}$ and $\gamma \beta = \frac{\sqrt{ Q^2}}{2 M}$,
which leads to
\begin{equation}
\label{app9}
(p'^{0}, p'^3)_{\rm TRF} =  \frac{1}{2 M}  \left( 2 M^2 + Q^2, \sqrt{Q^2(4 M^{2} + Q^2)} \right).
\end{equation}
In the TRF, both time and space components of the current $J^\mu$ in Eq.~(\ref{eq1}) are non-vanishing and explicitly they are 
given by
\bea\label{app10}
J_{\rm TRF}^{0}(q^{2})&=&\frac{4 M^{2} + Q^2}{2 M} F(q^{2}),
\nonumber\\
J_{\rm TRF}^{3}(q^{2})&=&\frac{\sqrt{Q^2} \sqrt{4 M^{2} +Q^2}}{2 M} F(q^{2}),
\end{eqnarray}
where $q^2 =(q^0)^2 - (q^3)^2 = -Q^2$. 
One should note, however, that the longitudinal charge density in TRF is not a static quantity but depends on time 
because of the fact that $q^0=\gamma\beta Q \neq 0$.

\subsection{Longitudinal charge density in LF coordinate space}

Recently, a general procedure was introduced to obtain frame-independent three-dimensional LF coordinate-space wave
functions~\cite{MB19}. 
In addition to the two dimensional transverse spatial variable ${\bf b}$ given by Eq.~(\ref{eq24c}), the longitudinal boost-invariant
dimensionless spatial variable $\tilde{z}=p^{+} x^{-}$ was also introduced.

In the present $(1+1)$-dimensional model calculations, the longitudinal charge density in LF coordinate space 
evaluated at $x^+=0$, as a Fourier transform of the form factor, can be defined by
\begin{equation}
\label{eqLF1}
\rho_{\rm LF}^{}(x^{-})=\frac{1}{4\pi}\int dq^{+} ~F(q^{2}) ~e^{\frac{i}{2} q^{+} x^{-}},
\end{equation}
where $\rho_{\rm LF}^{}(x^{-})$ satisfies $\int \rho_{\rm LF}^{}(x^{-})~ dx^{-}=1$. 
Using $q^{+}=\bar{\beta} p^{+}$ and Eq.~(\ref{eq5}), $\rho_{\rm LF}^{}$ can be rewritten in terms of 
$({\tilde z}, {\bar\beta})$ modulo the $p^+$ factor as
\begin{equation}
\label{eqLF2}
\rho_{\rm LF}(\tilde{z})=\frac{1}{4\pi}\int d\bar{\beta} ~F(\bar{\beta}) ~e^{\frac{i}{2} \bar{\beta} \tilde{z}},
\end{equation}
where $0\leq\bar{\beta}_{+}\leq\infty$ for ${\bar\beta}={\bar\beta}_+$ and
$-1\leq\bar{\beta}_{-}\leq0$ for ${\bar\beta}={\bar\beta}_-$, respectively. Since 
$\bar{\beta}_-$ has a closed range, which takes advantage over $\bar{\beta}_+$,
we obtain $\rho_{\rm LF}^{}(\tilde{z})$ integrating over $\bar{\beta}_-$ in Eq.~(\ref{eqLF2}).

\begin{figure}
\includegraphics[width=9cm, height=6cm]{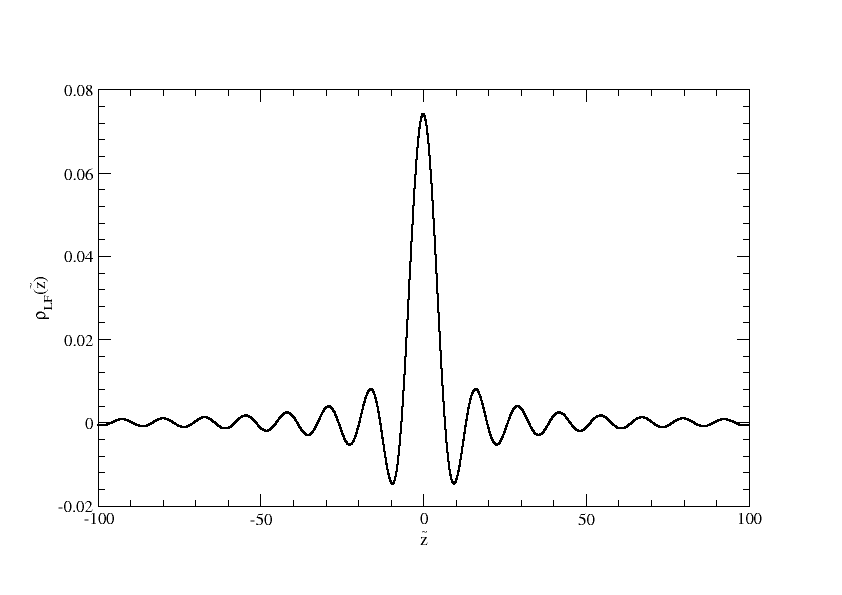}
\caption{\label{fig6} The longitudinal charge density for $\pi^+$ in LF coordinate space ${\tilde z}$.}
\end{figure}

Figure~\ref{fig6} shows the longitudinal charge density $\rho_{\rm LF}^{}(\tilde{z})$ for $\pi^+$ in a 
wide range of $-100\leq{\tilde{z}}\leq 100$. 
This shows that $\rho_{\rm LF}^{}(\tilde{z})$ has a very long and oscillating tail behavior of $\tilde{z}$, 
which appears consistent with the result shown in Ref.~\cite{MB19} for the case of two-constituents of a Fock-space component.
From the inverse Fourier transform of Eq.~(\ref{eqLF2}),
\begin{eqnarray}
\label{eqLF3}
F(\bar{\beta})&=&\int d\tilde{z}~ e^{-\frac{i}{2} \bar{\beta}\tilde{z}} \rho_{\rm LF}^{} (\tilde{z})\nonumber\\
&=&1-\frac{\bar{\beta}^{2}}{8}\langle \tilde{z}^2 \rangle+\cdots,
\end{eqnarray}
one may obtain the mean square charge radius in ${\tilde z}$ as
$\langle \tilde{z}^2 \rangle=-8 ~\partial F/ \partial \bar{\beta}^{2}\vert_{\bar{\beta}^{2}=0}$ where
$\langle \tilde{z}^2 \rangle=\int d\tilde{z}~ \tilde{z}^2 \rho_{\rm LF}^{} (\tilde{z})$.

\section{Numerical results}
\label{sec:4}

In our numerical calculations, we analyze scalar $\pi$, $K$, and $D$ meson form factors.  
For these analyses, we use  the constituent quark and antiquark masses as $m_u=m_d = 0.25~\mbox{GeV}$,
$m_s = 0.5~\mbox{GeV}$, and $m_c = 1.8~\mbox{GeV}$ as in Ref.~\cite{CJ99}. 
The used physical meson masses are 
$M_{\pi^{\pm}}= 0.14~\mbox{GeV}$, 
$M_{K^{\pm}}= 0.494~\mbox{GeV}$,  $M_{K^0}= 0.497~\mbox{GeV}$,
$M_{D^{\pm}}= 1.870~ \mbox{GeV}$, and 
$M_{D^0}= 1.865~ \mbox{GeV}$, respectively.
It should be noted from our constituent masses that $M^2 < m^2_q + m^2_{\bar Q}$ for $\pi$ and $K$  but
$M^2 > m^2_q + m^2_{\bar Q}$ for  $D$ meson case, while all the mesons satisfy the bound state condition,
$M< m_q + m_{\bar Q}$. 
This means that $\pi$ and $K$ are strongly bound states but $D$ is weakly bound state of which properties will be
discussed in the following numerical calculations.

In the previous work for the $\phi^3$ model in $(3+1)$ dimensions~\cite{CJ99}, two of us analyzed the form factors 
in three different reference frames, namely, (i) purely longitudinal ($q^+\neq 0$ and ${\bf q}_\perp=0$) frame defined in 
the timelike $q^2>0$ region, (ii) purely longitudinal ($q^+\neq 0$ and ${\bf q}_\perp=0$) frame defined in the spacelike $
q^2<0$ region, and (iii) the ($q^+=0$ with $q^2 = -{\bf q}^2_\perp$) frame,%
\footnote{The details can be found in Eqs. (10), (19), and (23) of Ref.~\cite{CJ99}.}
to confirm that all of three reference frames give exactly the same numerical results for the form factor in the entire $q^2$ region.
So, when we refer the ``direct results" from $(3+1)$ dimensions, we mean the results obtained from any of those three results
in Ref.~\cite{CJ99}.
Likewise, the direct results from $(1+1)$ dimensions indicate those obtained by using Eq.~(\ref{eq9}) or (\ref{eq16}) in the present work. 
On the other hand, ``DR results" refer to those obtained from the dispersion relations given by Eq.~({\ref{eq18}).
Comparing the two results obtained from both $(1+1)$- and $(3+1)$-dimensions, we shall also estimate the effects of
the transverse momenta ${\bf k}_\perp$ of the quark and anitquark on the EM form factors.

Shown in Fig.~\ref{fig7} is the profile of the intrinsic longitudinal charge density $\rho_{\pi}(Q,r_z)$ for $\pi^+$ and its contour plot 
in phase space $(Q, r_z)$ of $-10\leq Q\leq 10~\mbox{GeV}$ and $-1\leq r_z \leq 1~\mbox{fm}$. 
The momentum-dependent $\rho(Q,r_z)$ is defined as 
$\rho(r_z) =\frac{1}{2\pi}\int F(Q^2) e^{-i Q r_z} \, dQ \equiv \int \rho(Q, r_z) \, dQ$. 
We also show in Fig.~\ref{fig8} the profiles of the intrinsic longitudinal charge densities for $(K^+, K^0)$ and $(D^+, D^0)$ 
in phase space of $-10\leq Q\leq 10~\mbox{GeV}$ and $-1\leq r_z \leq 1~\mbox{fm}$. 
These figures confirm that $\rho(Q,r)$ is symmetric under $Q\to -Q$ and $r_z \to -r_z$ as expected.
The generic structures of $\rho(Q,r_z)$ for charged particles ($\pi^+, K^+, D^+$) look similar to each other. 
Likewise, the generic structures of neutral particles ($K^0, D^0$) are similar to each other. 
On the other hand, the density profiles of charged particles are quite different from those of neutral ones.

\begin{figure}[t]
\includegraphics[width=0.45\columnwidth]{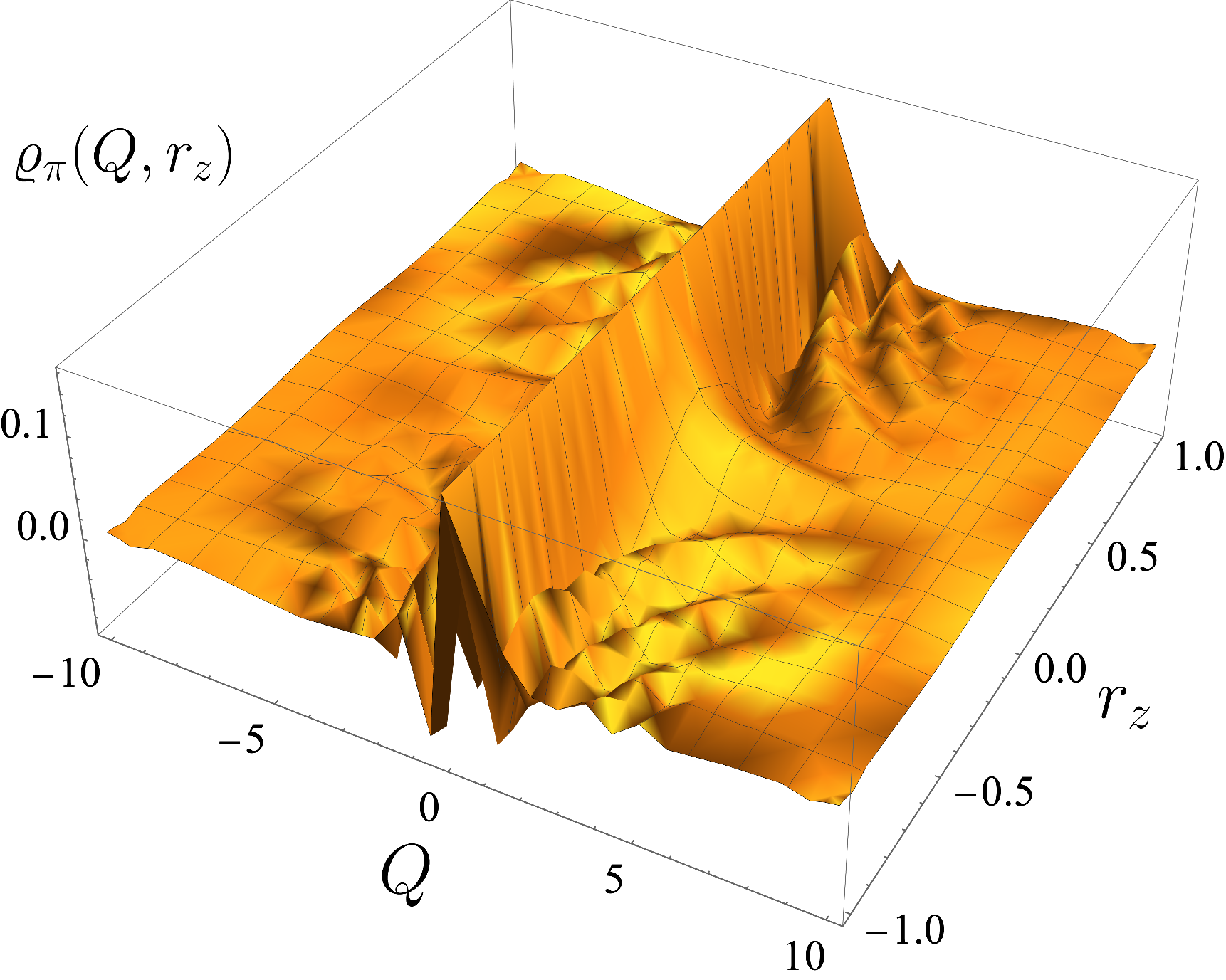}
\includegraphics[width=0.45\columnwidth]{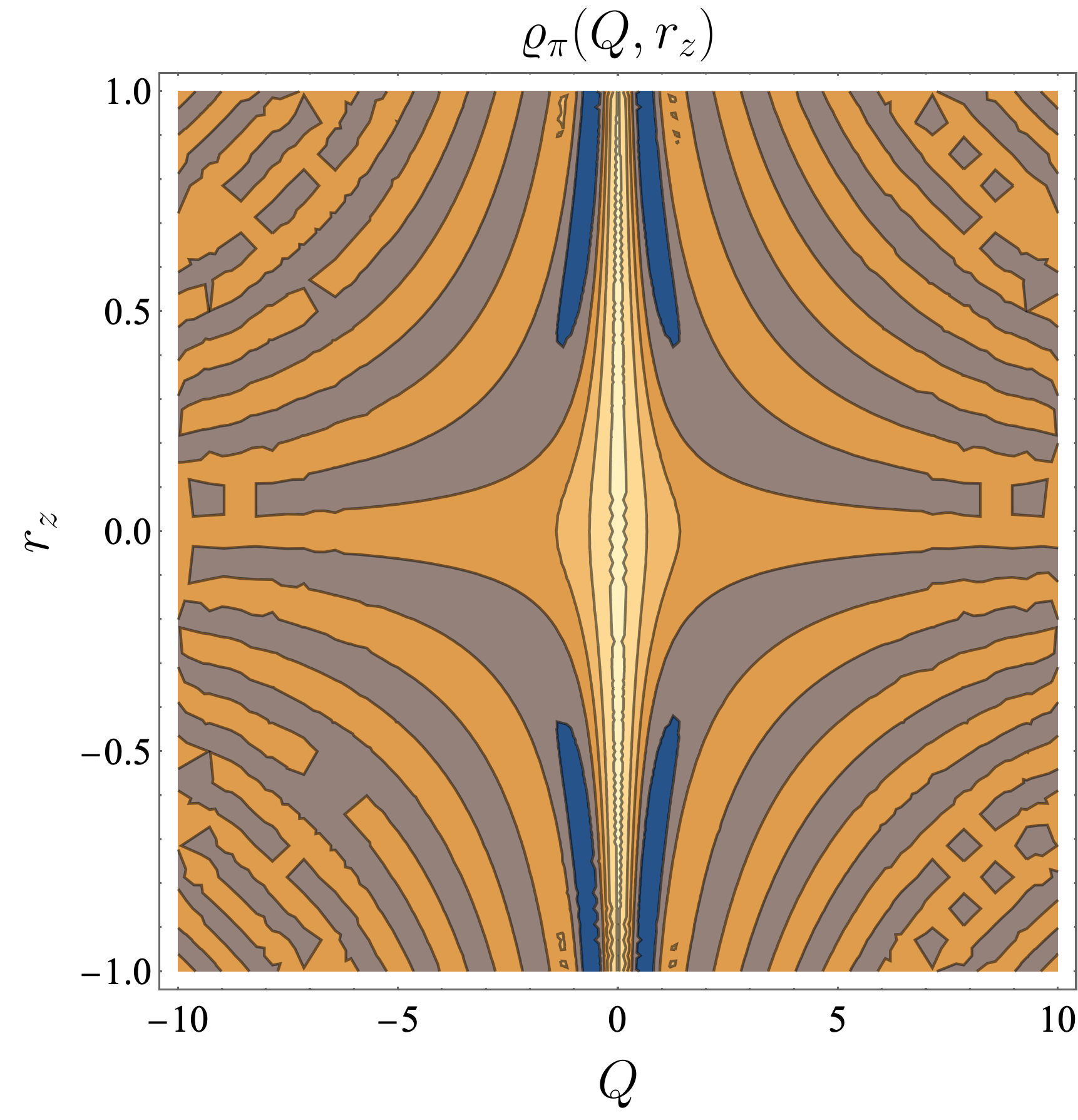}
\caption{\label{fig7} Profile of the intrinsic longitudinal charge density $\rho_{\rm ILD}^{}(Q,r_z)$ for $\pi^+$ 
and its contour plot  in phase space $(Q,r_z)$ of $-10\leq Q\leq 10$~GeV and $-1\leq r_z \leq 1$~fm.}
\end{figure}

\begin{figure}[t]
\includegraphics[width=4.2cm, height=4.2cm]{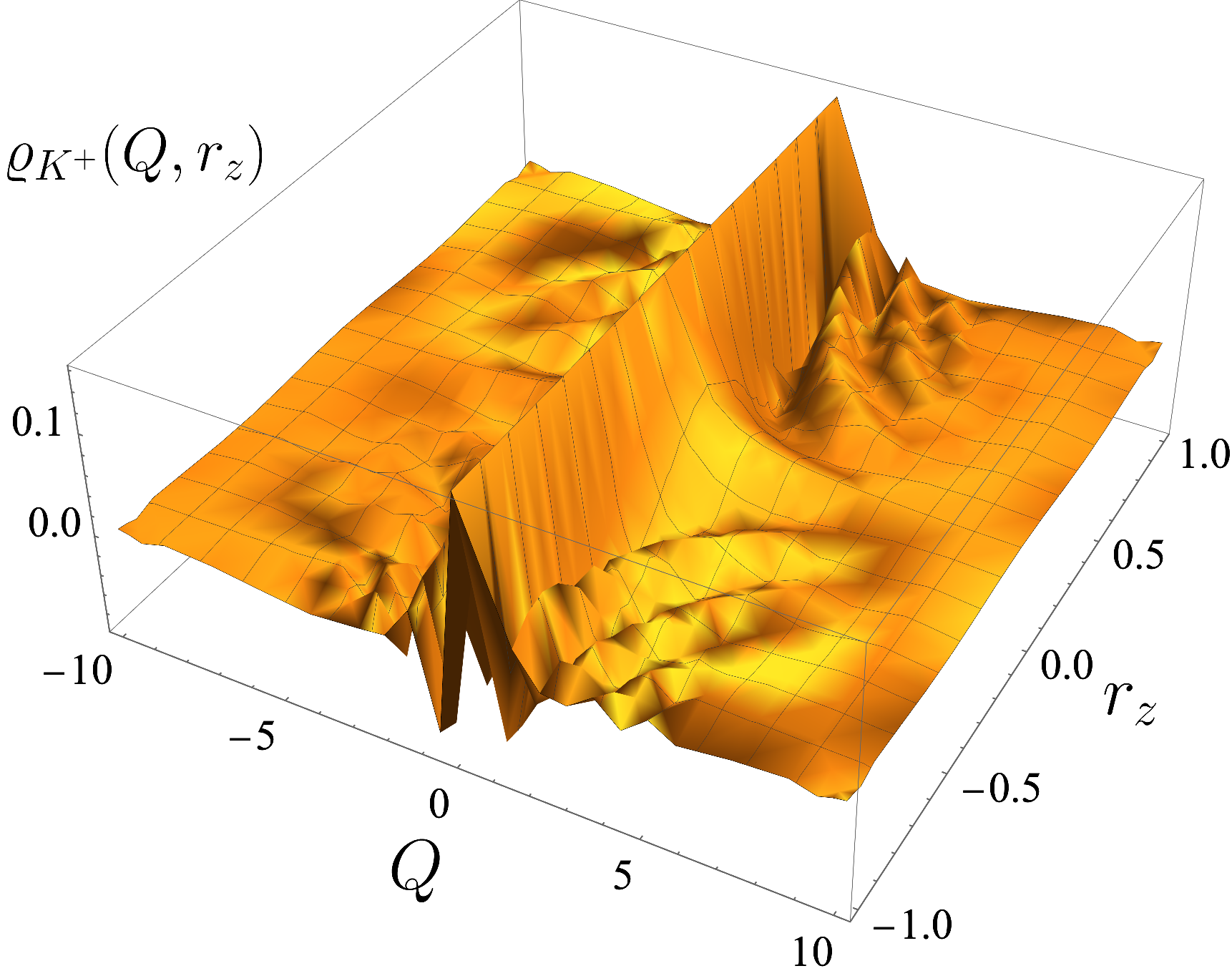}
\includegraphics[width=4.2cm, height=4.2cm]{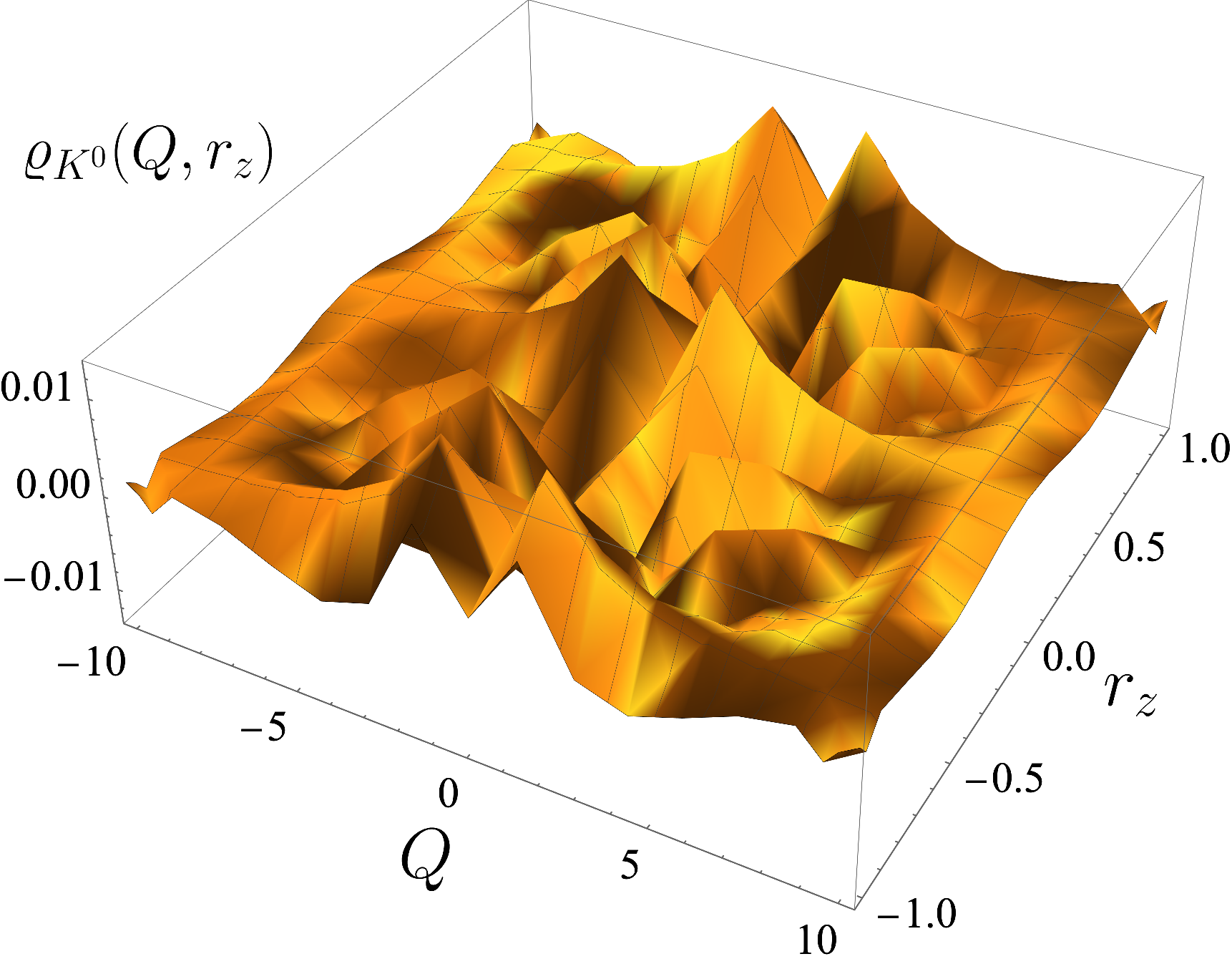}
\includegraphics[width=4.2cm, height=4.2cm]{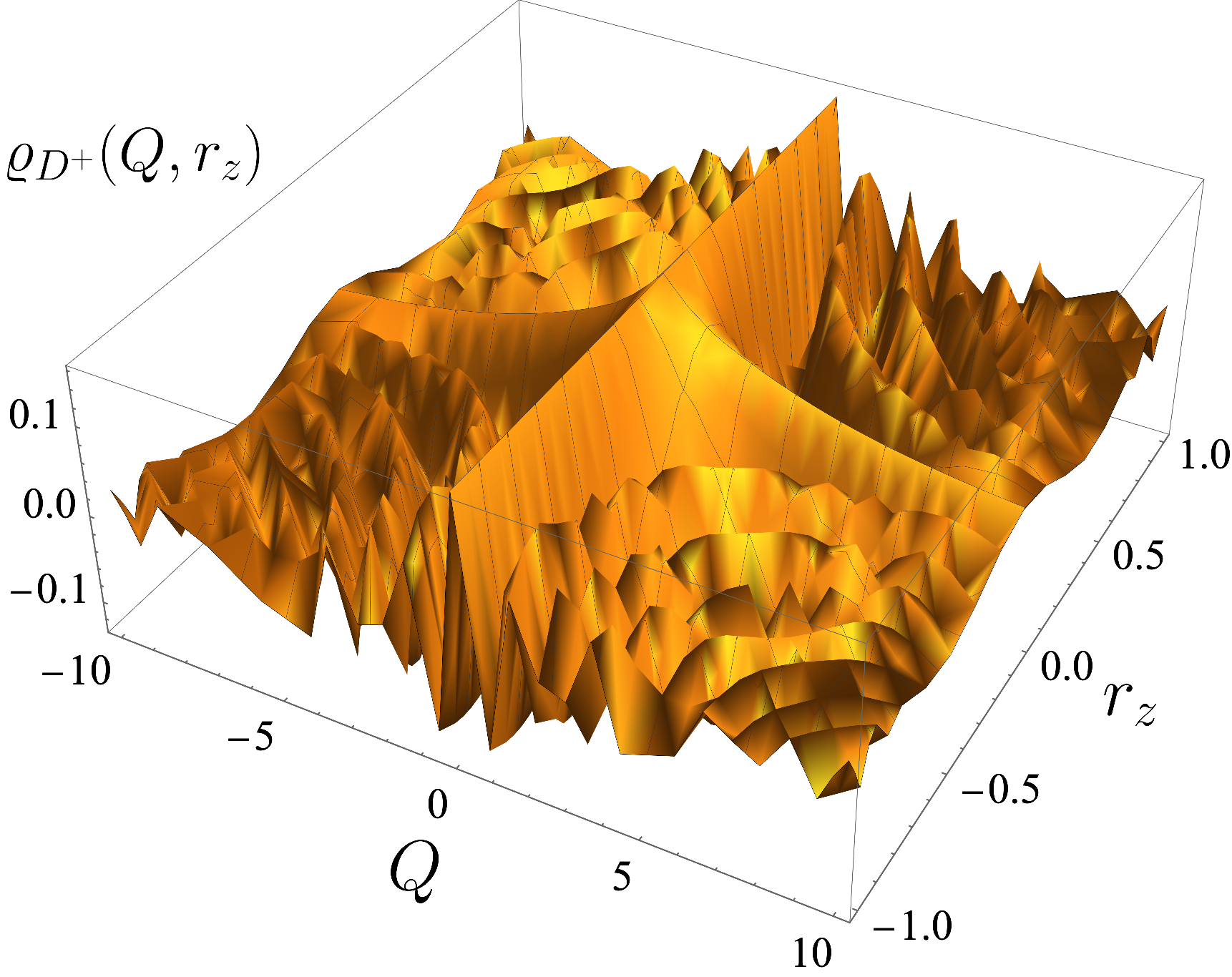}
\includegraphics[width=4.2cm, height=4.2cm]{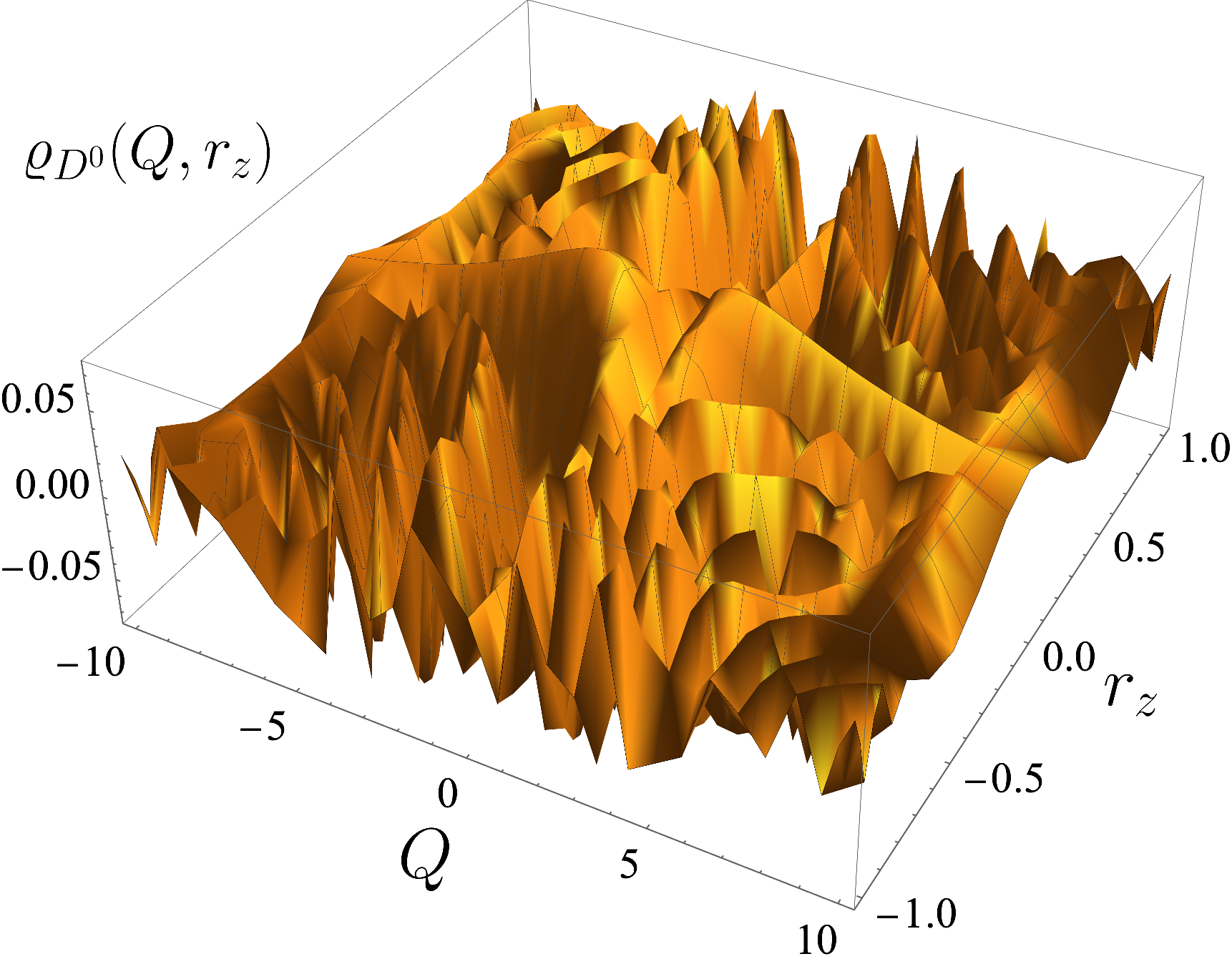}
\caption{\label{fig8} Profiles of the intrinsic longitudinal charge densities $\rho_{\rm ILD}(Q,r_z)$ for $(K^+, K^0)$ and $(D^+, D^0)$ in phase space
of $-10\leq Q\leq 10$ GeV, $-1\leq r_z \leq 1$ fm, respectively. }
\end{figure}

We present the intrinsic longitudinal charge densities $\rho(r_z)$ for charged $(\pi^+, K^+, D^+)$ mesons in Fig.~\ref{fig9}(a) and
for neutral $(K^0, D^0)$ mesons in Fig.~\ref{fig9}(b), respectively. 
The gap between $\rho_{\pi^+}(r_z)$ (solid line) and  $\rho_{K^+}(r_z)$ (dotted line) is very small and it is
hard to distinguish them in Fig.~\ref{fig9}(a).
Compared to the light $(\pi, K)$ mesons, the charge density of heavy $D$ meson is narrowly peaked around $r_z=0$.
Figure~\ref{fig9}(b) also shows the charge density behavior of neutral $(K^0, D^0)$ mesons which satisfy $\int \rho(r_z) dr_z =0$.

\begin{figure*}[t]
\includegraphics[width=7cm, height=7cm]{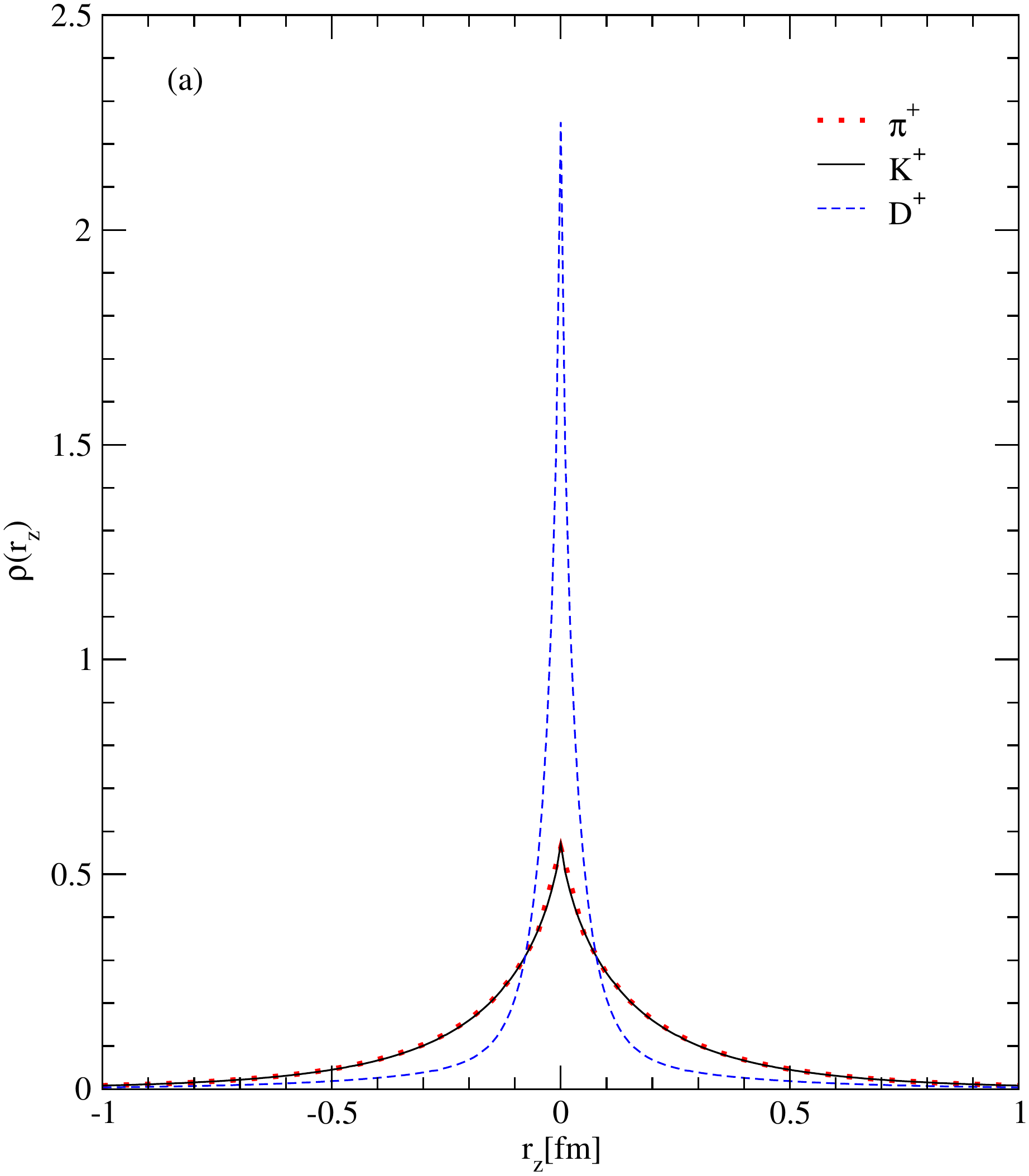}
\includegraphics[width=7cm, height=7cm]{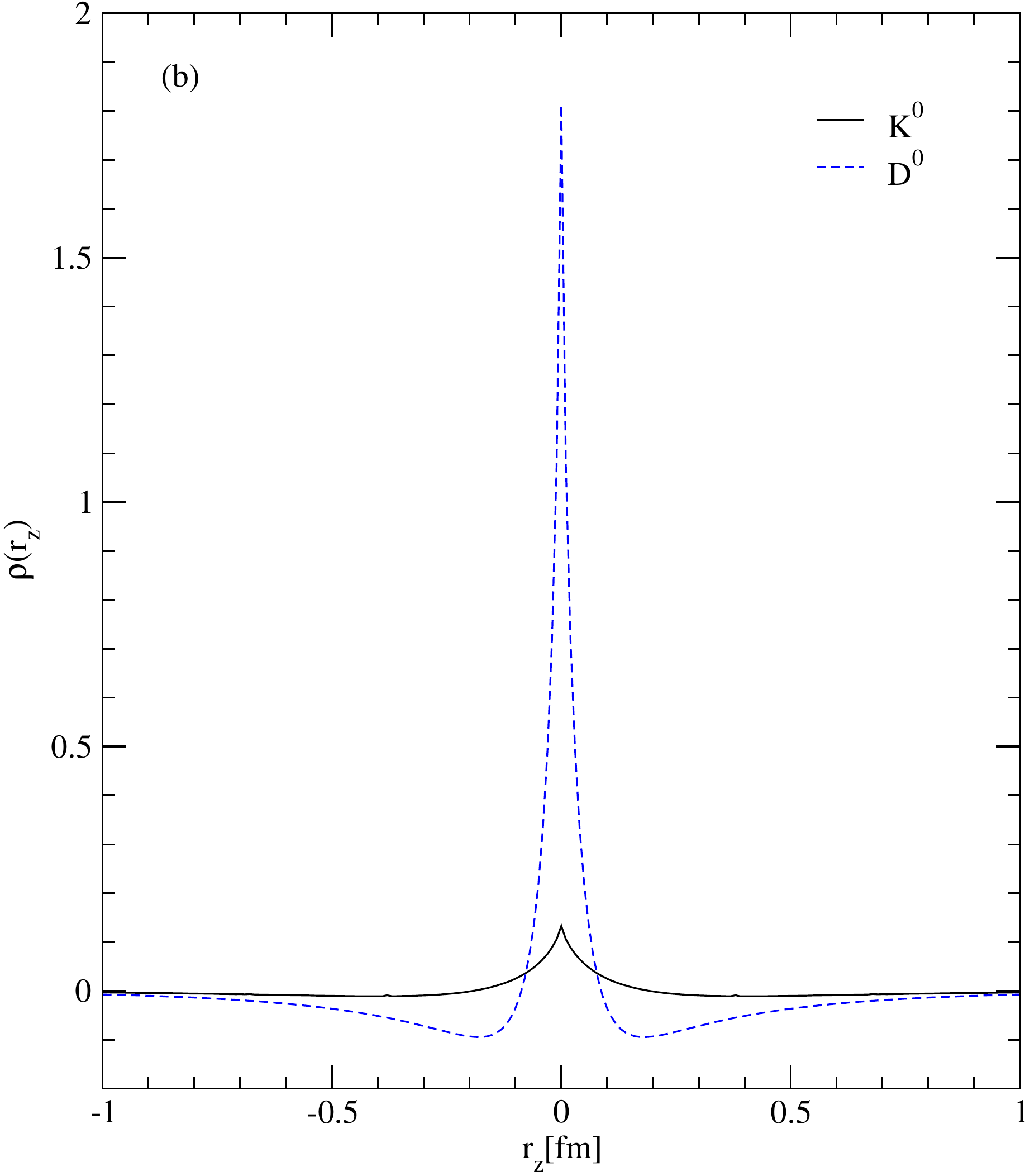}
\caption{\label{fig9} The intrinsic longitudinal charge densities $\rho_{\rm ILD}^{}(r_z)$ of (a) $(\pi^+, K^+, D^+)$ and (b) $(K^0, D^0)$
for $-1\leq r_z \leq 1$~fm.}
\end{figure*}

\begin{figure*}[t]
\includegraphics[width=7cm, height=7cm]{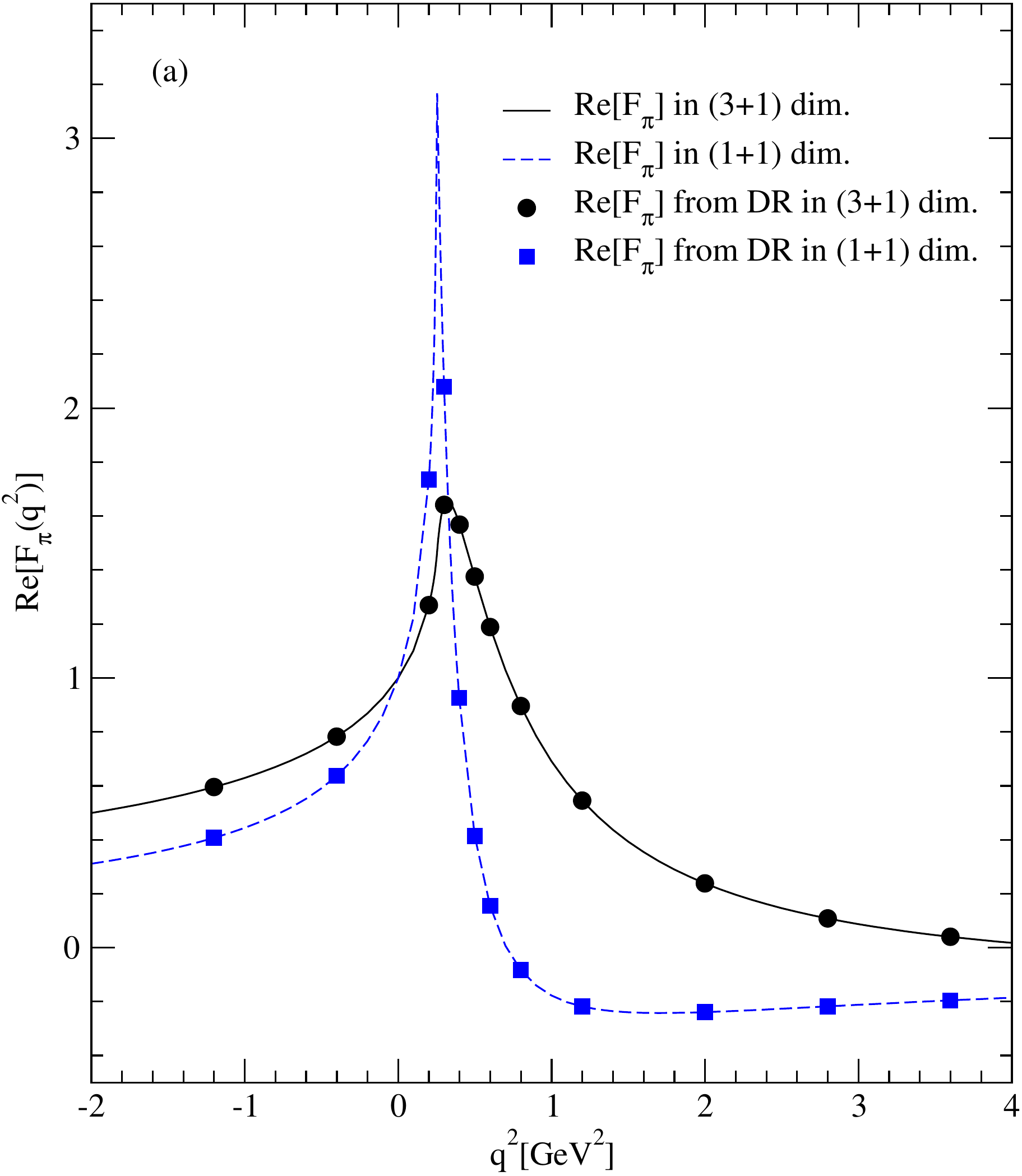}
\includegraphics[width=7cm, height=7cm]{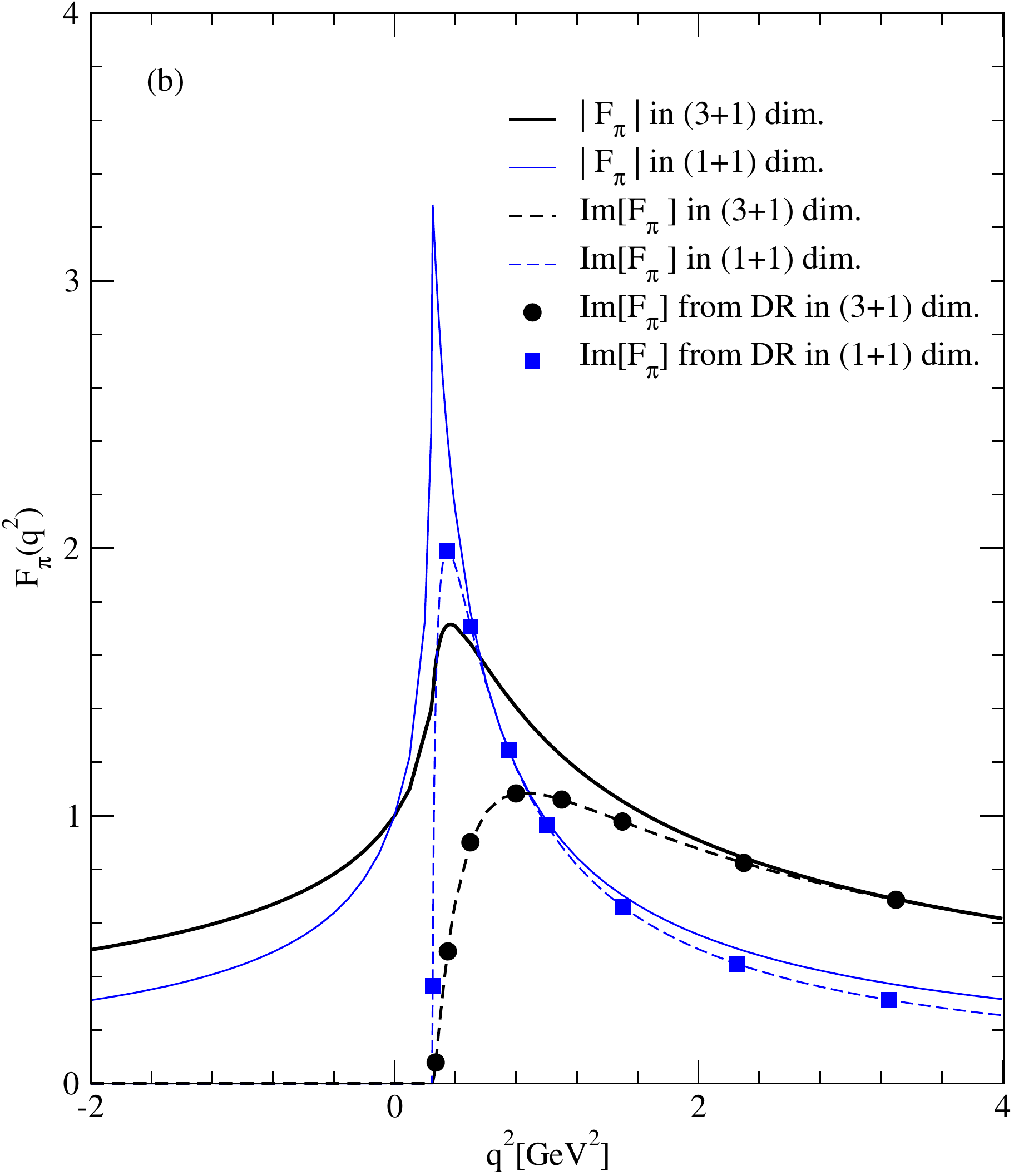}
\caption{\label{fig10}  EM form factor of the pion in $(1+1)$- and $(3+1)$-dimensions: 
(a) Re[$F_\pi(q^2)$] and (b) Im[$F_\pi(q^2)$] and $|F_\pi(q^2)|$ for $-2\leq q^2\leq 4$~GeV$^2$ compared with 
the results obtained from the dispersion relations. }
\end{figure*}

In Fig.~\ref{fig10}, we show the EM form factor of the pion obtained in $(1+1)$- and $(3+1)$-dimensions for $-2\leq q^2\leq 4~\mbox{GeV}^2$.
The black and blue lines represent the direct results obtained from the form factors in $(3+1)$- and $(1+1)$-dimensions, respectively. 
The corresponding $(3+1)$- and $(1+1)$-dimensional results obtained from the dispersion relations are denoted by black circles 
and blue squares. 
Figure~\ref{fig10}(a) represents Re[$F_\pi(q^2)$], and Fig.~\ref{fig10}(b) includes both Im[$F_\pi(q^2)$] and  $|F_\pi(q^2)|$. 
Close inspection of the figures leads to the following comments.
Firstly, our ``direct results" for both Re[$F_\pi(q^2)$] and Im[$F_\pi(q^2)$] in $(1+1)$- and  $(3+1)$-dimensions show 
complete agreement with the ``DR results" in corresponding dimensions, respectively.
Secondly, the imaginary parts (dashed lines) of the form factors in Fig.~\ref{fig10}(b) obtained from both $(1+1)$- and $(3+1)$-dimensions
start at the normal threshold $q^2_{\rm min}=4 m^2_{u(d)}=0.25~\mbox{GeV}^2$, which is consistent with the condition for 
$M^2 < m^2_q + m^2_{\bar Q}$ case. 
For high $q^2$ region, the imaginary parts of the form factors are shown to dominate over the real part.
Thirdly, the total form factors $|F_\pi (q^2)|$ (solid lines) in both $(1+1)$- and $(3+1)$-dimensions produce a $\rho$ meson-type peaks
consistent with the vector meson dominance (VMD). 
However, we do not claim that this model indeed reproduces all the features of the VMD phenomena since more realistic 
phenomenological models may have to incorporate more complex mechanisms such as the initial- and final- state interactions. 
Finally, in that the difference between the two results in $(1+1)$- and $(3+1)$-dimensions measures the effects of transverse momenta 
of the constituents, one can see that the effects of ${\bf k}_\perp$ reduce the charge radii and broaden the widths of the peaks. 
Therefore, while the qualitative behaviors of the form factors in both dimensions are not much different from each other, 
their quantitative behaviors are quite sizable due to the effects of the transverse momenta of the constituents.

\begin{figure*}
\includegraphics[width=5cm, height=5cm]{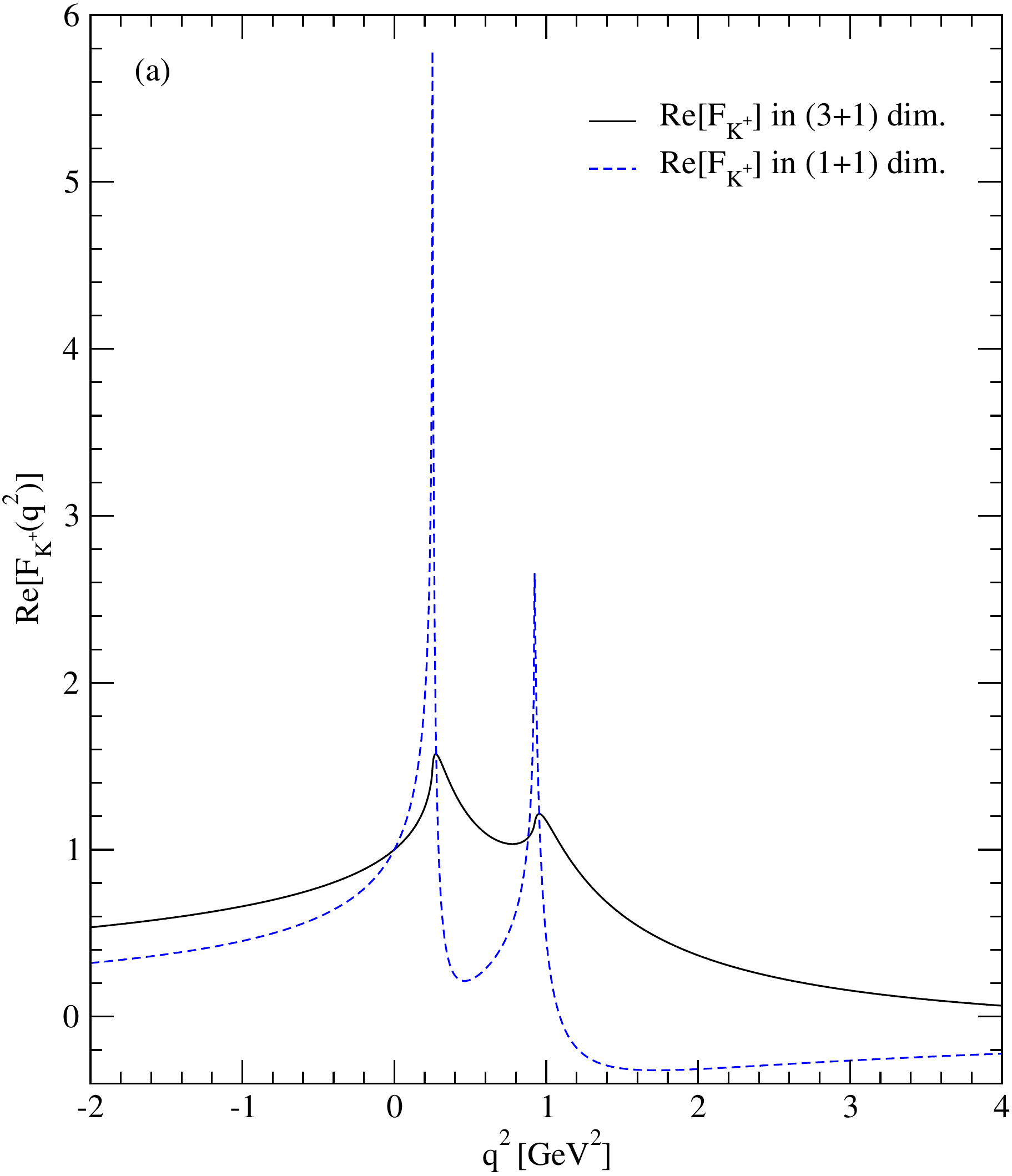}
\includegraphics[width=5cm, height=5cm]{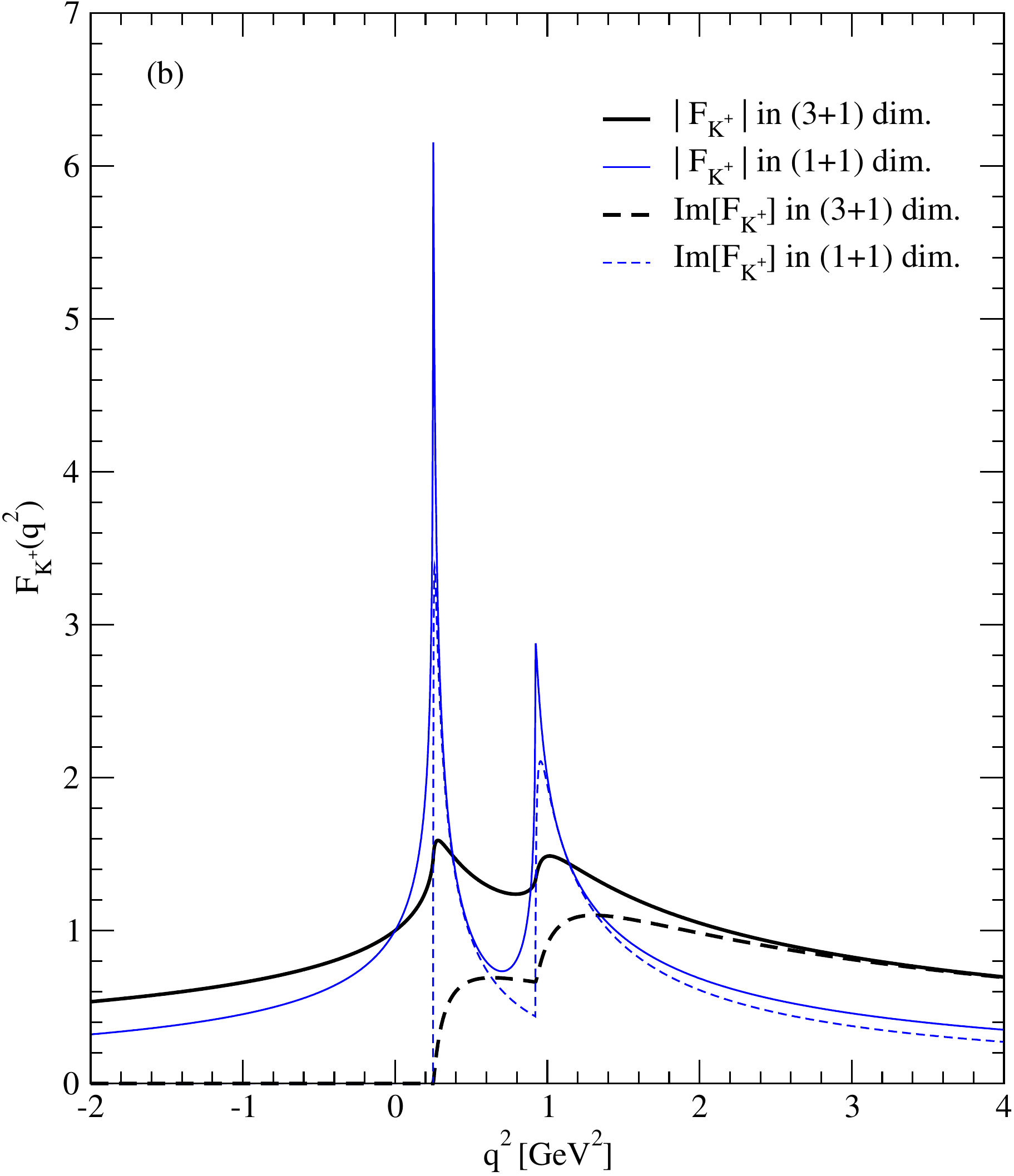}
\\
\includegraphics[width=5cm, height=5cm]{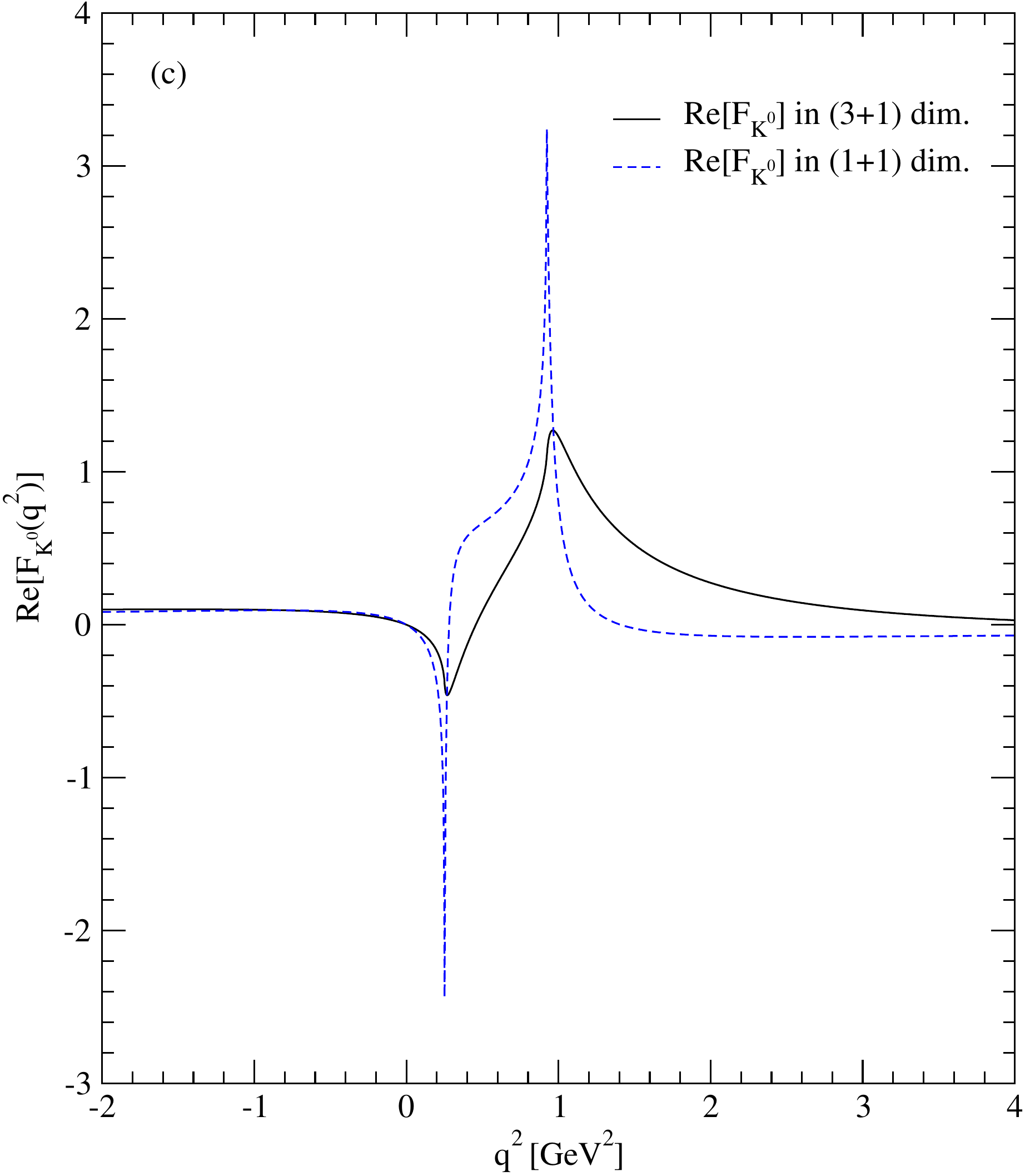}
\includegraphics[width=5cm, height=5cm]{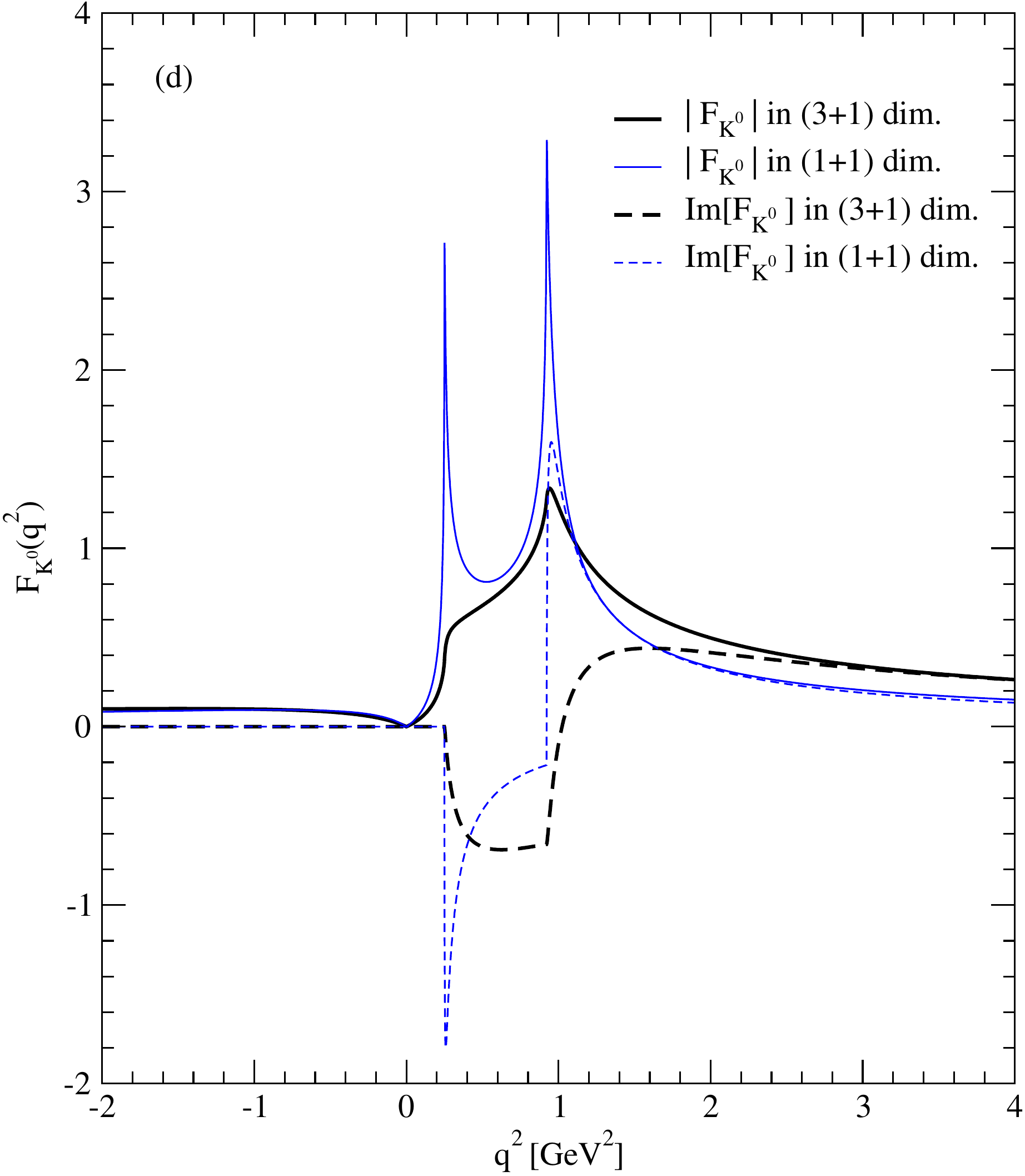}
\caption{\label{fig11}  EM form factors of $K^+$ and $K^0$ mesons in $(1+1)$- and $(3+1)$-dimensions: 
(a) Re[$F_{K^+}(q^2)$],  (b) Im[$F_{K^+}(q^2)$] and $|F_{K^+}(q^2)|$,
(c) Re[$F_{K^0}(q^2)$],  and (d) Im[$F_{K^0}(q^2)$] and $|F_{K^0}(q^2)|$
for $-2\leq q^2\leq 4$~GeV$^2$. }
\end{figure*}

Figure~\ref{fig11} shows the EM form factors of $K^+(u\bar{s})$ and $K^0(d\bar{s})$ mesons obtained from $(1+1)$- and $(3+1)$-dimensions
for $-2\leq q^2\leq 4~\mbox{GeV}^2$. 
The same line codes are used as in Fig.~\ref{fig10}. 
The direct results and the DR results for the unequal quark mass cases such as $K$ and $D$ mesons coincide and we do not explicitly 
display the DR results in Fig.~\ref{fig11}. 
As in the case of the pion, both $K^+$ and $K^0$ have the normal singularities.
However, $K$ mesons have two thresholds, namely, one at $q^2_{\rm min}=4 m^2_{u}$ (or $4 m^2_d$) and the other at $q^2_{\rm min}=4 m^2_{s}$.
While we have in principle two vector-meson-type peak (i.e. $\rho$ and $\phi$), one can see in Fig.~\ref{fig11} that only the $\phi$ meson-type 
peak would be observable for the timelike kaon EM form factors above the physical threshold at $q^2_{\rm min}= 4M^2_{K^{+(0)}}$ 
as the $\rho$ meson-type peak is kinematically below the physical threshold, i.e. $4 m^2_{u(d)} <  4M^2_{K^{+(0)}}$.
Again, the differences between the $(1+1)$- and $(3+1)$-dimensional results reside in the effects of transverse momenta of the constituents
which play the role of broadening the widths and flatten the heights of the form factors.
The sign flips for both Re[$F_{K^0}(q^2)$] and Im[$F_{K^0}(q^2)$] between the two peaks come from the different sign of electric charges of $d$ and $\bar{s}$
quarks. 
One can find that the $K^0$ meson has the primary and secondary peaks near the heavy $s$ and light  $d$ quark
threshold $q^2_{\rm min}=4 m^2_{s(d)}$, respectively, while it is the opposite for the $K^+$ meson. 
The spacelike $q^2$ region of $|F_{K^{+,0}}(q^2)|$ in Figs.~\ref{fig11}(b,d) shows that $K^+$ has a positive mean-square charge radius 
while $K^0$ has a negative one.

\begin{figure*}
\includegraphics[width=5cm, height=5cm]{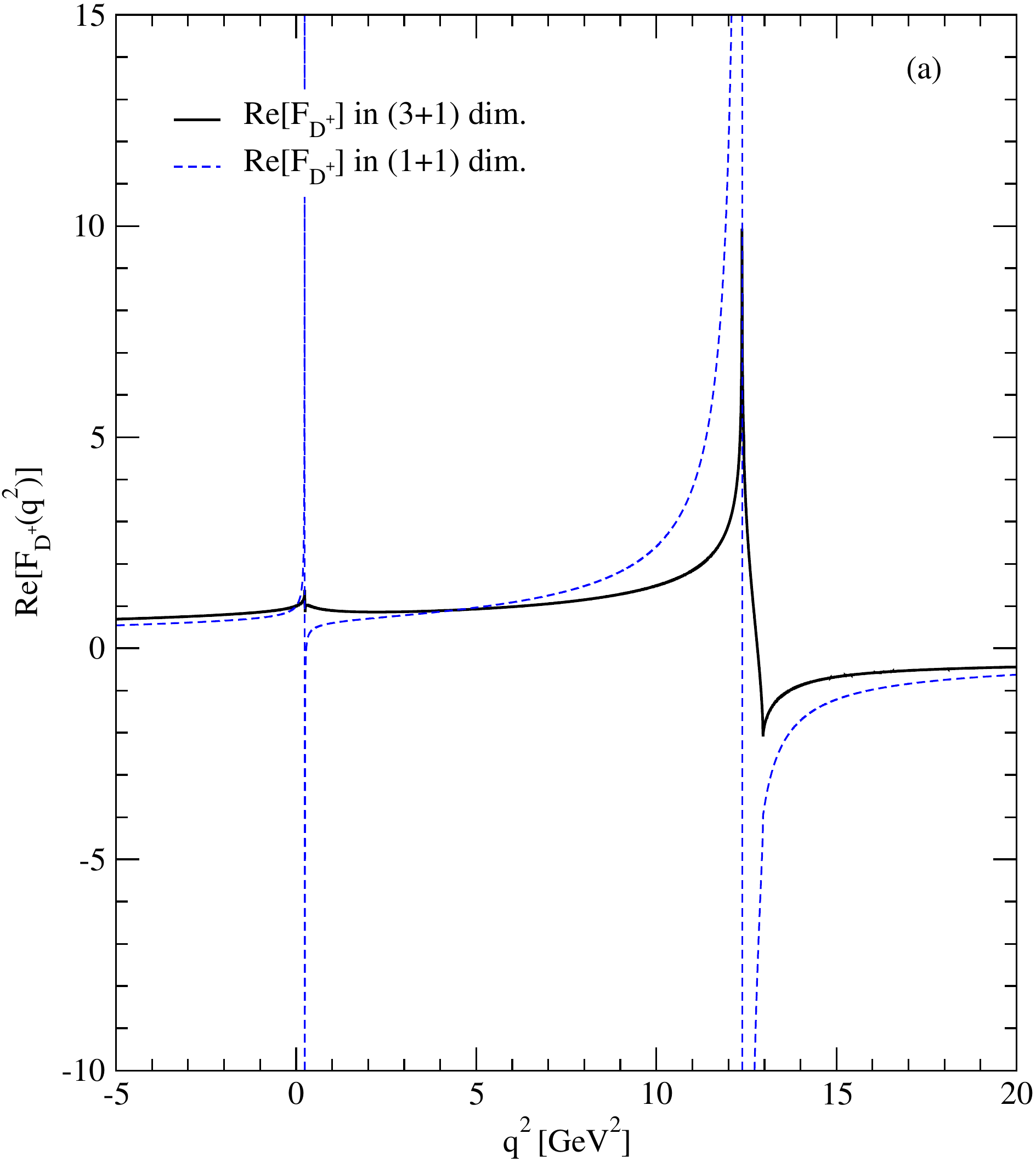}
\includegraphics[width=5cm, height=5cm]{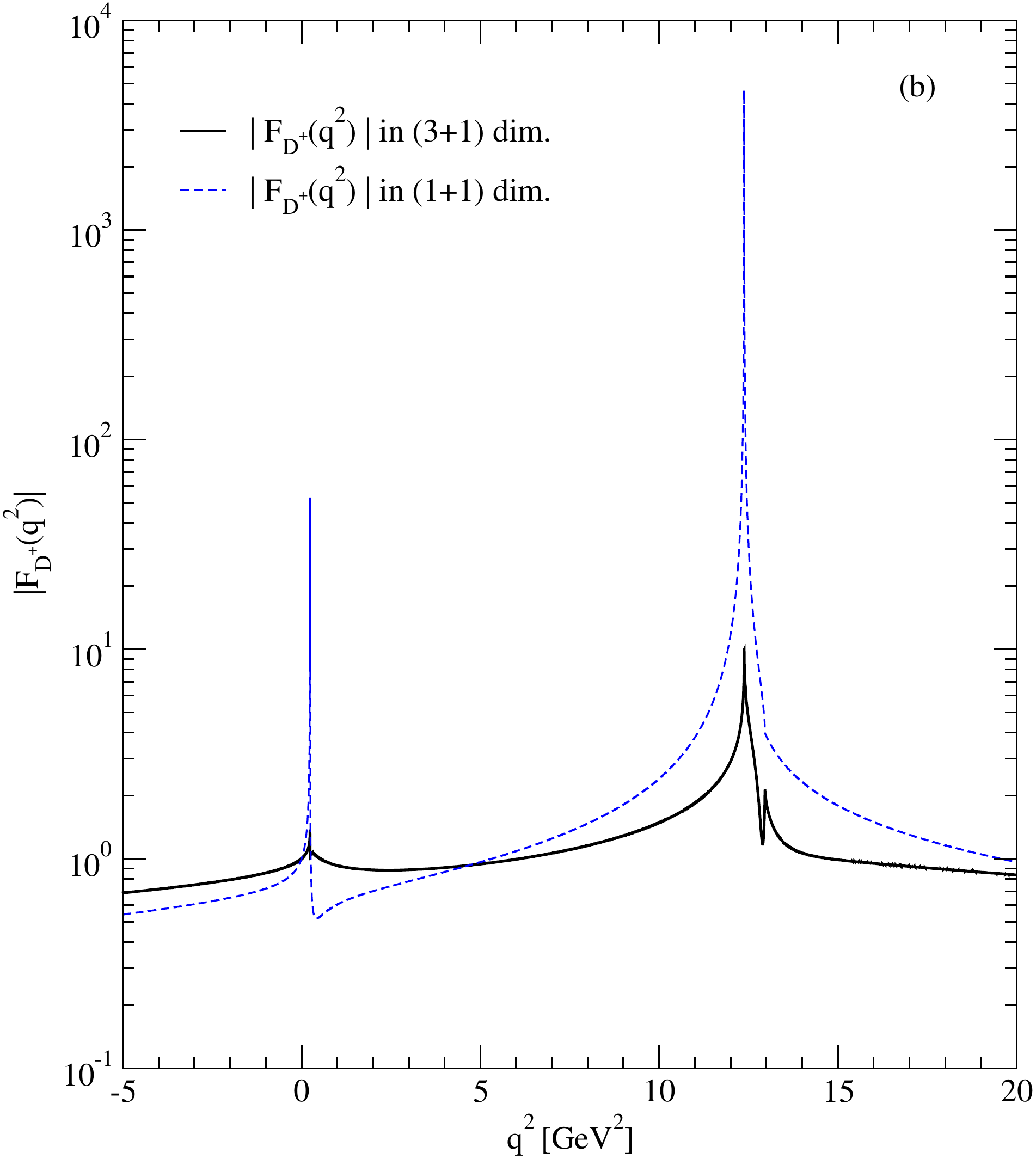}
\\
\includegraphics[width=5cm, height=5cm]{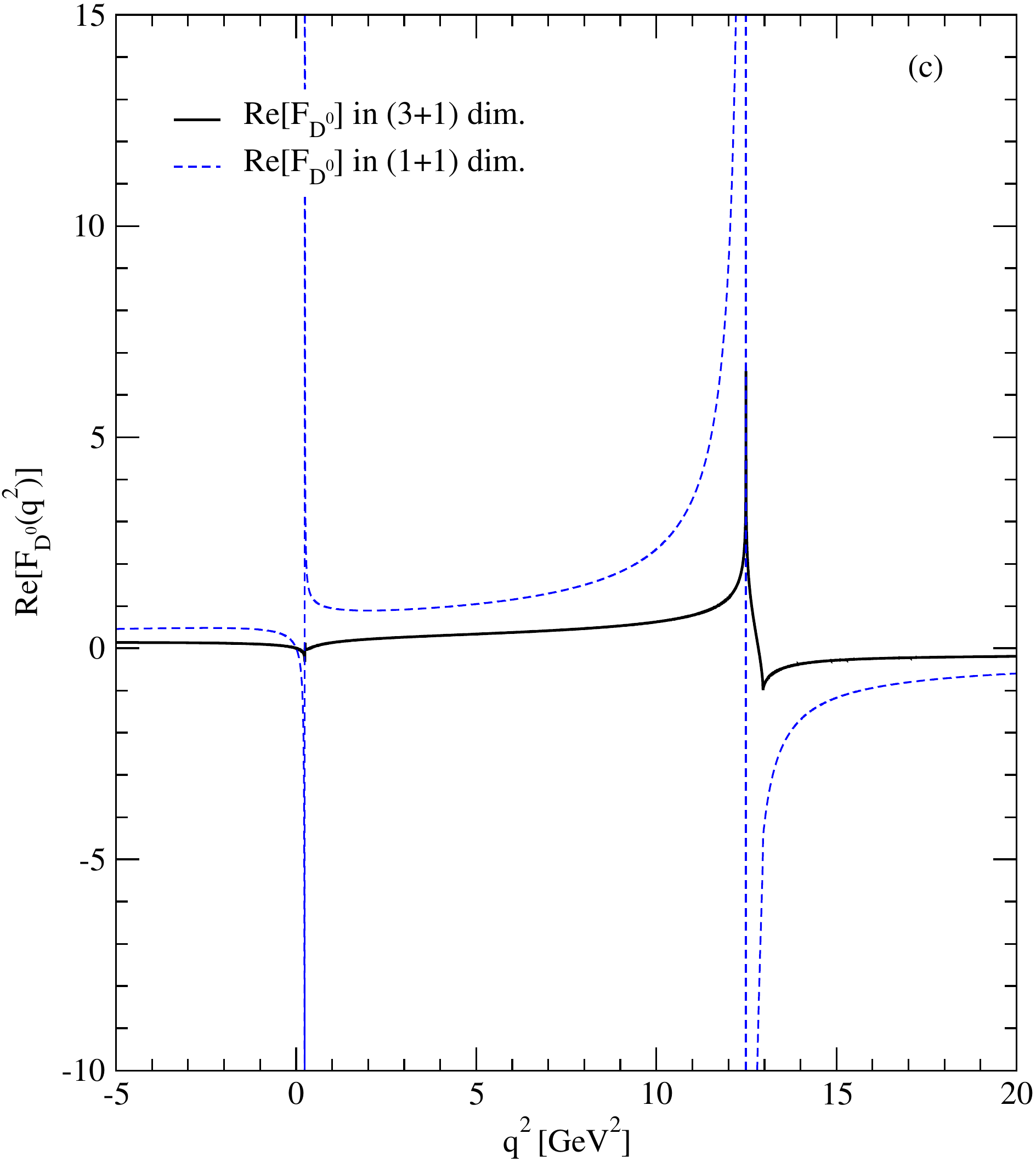}
\includegraphics[width=5cm, height=5cm]{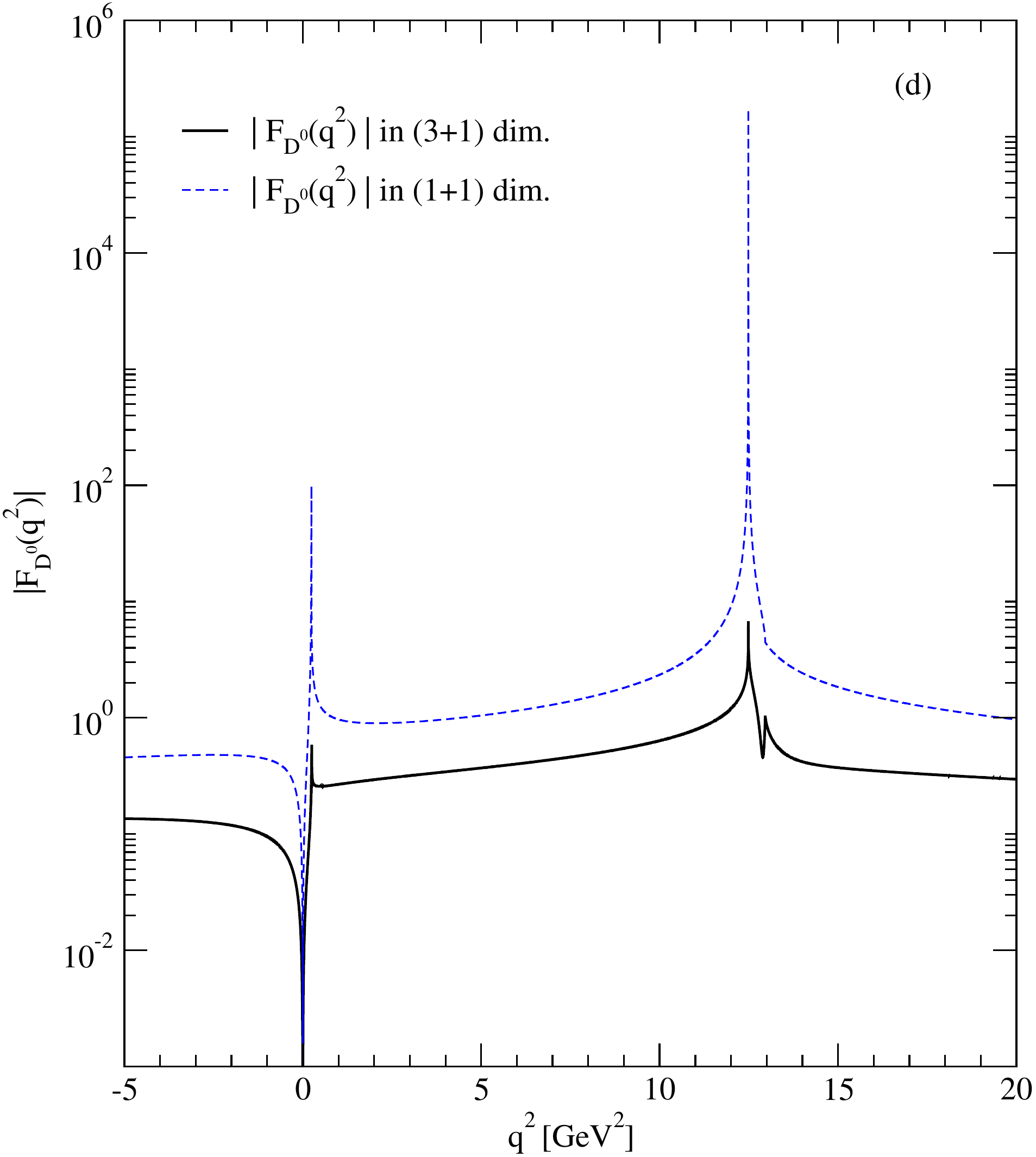}
\caption{\label{fig12}  EM form factors of $D^+$ and $D^0$ mesons in $(1+1)$- and $(3+1)$-dimensions: 
(a) Re[$F_{D^+}(q^2)$],  (b) Im[$F_{D^+}(q^2)$]   and  $|F_{D^+}(q^2)|$,
(c) Re[$F_{D^0}(q^2)$],  and (d) Im[$F_{D^0}(q^2)$]   and  $|F_{D^0}(q^2)|$
for $-5\leq q^2\leq 20$  GeV$^2$. }
\end{figure*}

We present the EM form factors of $D^+(c\bar{d})$ and $D^0(c\bar{u})$ mesons obtained in $(1+1)$- and $(3+1)$-dimensions 
for $-5\leq q^2\leq 20~\mbox{GeV}^2$ in Fig.~\ref{fig12}. 
The same line codes are used as in Fig.~\ref{fig10}. 
Although the generic features of $D$ meson form factors are similar to the case of $K$ mesons, several comments are in order.
As explained before, the weakly bound state such as the $D$ meson in the present model calculation has anomalous thresholds.
Both $(1+1)$-and $(3+1)$-dimensional results of $|F_{D^{+(0)}}(q^2)|$ in Figs.~\ref{fig12}(b,d) indeed show the presence of anomalous thresholds 
as given in Eq.~(\ref{eq19}), i.e., $q^2_{\rm min}\sim 0.24$~GeV$^2$ (compared to $4 m_{d(u)}=0.25$~GeV$^2$ for normal case) and
$q^2_{\rm min}\sim 12.4$~GeV$^2$ (compared to $4 m_{c}=12.96$~GeV$^2$ for normal case) for the
$\gamma^*$-${\bar{d}}(\bar{u})$ and $\gamma^*$-$c$ vertices, respectively. 
Both anomalous thresholds, however, appear just before the normal thresholds although the normal thresholds for the light quark sector 
are hard to be seen in Fig.~\ref{fig12}. 
Similar to the $K$ meson case, the EM form factors of both charged and neutral $D$ mesons have two unphysical peaks, i.e., $\rho$ and 
$J/\psi(1S)$ meson type peaks due to ${\bar d}$ (or ${\bar u}$) and $c$ quarks, respectively. 
However, the timelike form factors of $D$ mesons have no pole structures  for the physical 
 $q^2\geq 4 M_{D}^2$ region.
Finally, unlike the kaon case, the primary peaks for both $D^+$ and $D^0$ appear near the thresholds due to the heavy $c$ quark.
Figures~\ref{fig12}(b,d) also shows that $D^+$ has a positive mean-square charge radius while $D^0$ has a negative one.

\section{Summary and Conclusion}
\label{sec:5}

In the present work, we presented $(1+1)$-dimensional analysis of EM form factors of scalar mesons 
both for spacelike and timelike region in the solvable model, obtaining the analytic results of the one-loop triangle diagram both 
for the valence and non-valence contributions.  
Since the $q^+ \neq 0$ frame should be used in $(1+1)$ dimensions, it is inevitable to encounter the non-valence diagram 
arising from the particle-antiparticle pair creation (the so-called ``Z-graph").
While the valence contribution dominates for small $Q^2$ region, its role is taken over by the non-valence contribution 
as $Q^2$ gets larger indicating significant contributions from the higher-Fock components.
The leading asymptotic behavior of the form factor at high $Q^2$ in $(3+1)$ dimensions~\cite{Miller09}
has one more logarithmic power than our result $F^{S}_{\cal M}(Q^2\to\infty)\sim \ln Q^2/Q^2$
in $(1+1)$ dimensions, which is ascribed to the effects of the transverse momentum.
Our analytic results both in the spacelike region and the timelike region confirmed 
the analytic continuation from $Q^2$ in the spacelike region to $-Q^2$ in the timelike region.
In the timelike form factor given by Eq.~(\ref{eq15}), the imaginary part of $I^{q(Q)}_{T}(q^2)$ starts to develop at the anomalousl
threshold given by Eq.~(\ref{eq19}) for the weakly bound state with $M< m_q + m_{\bar Q}$ but $M^2 > m^2_q + m^2_{\bar Q}$, 
while it starts at the normal threshold $q^2 \geq 4 m^2_{q(Q)}$ for the strongly bound state with $M< m_q + m_{\bar Q}$ and 
$M^2 < m^2_q + m^2_{\bar Q}$.
We confirm that the DRs given by Eq.~(\ref{eq18}) are satisfied by the strongly bound state as well as by the weakly bound state. 
In particular, we note the importance of taking into account the infinitesimal width to remedy the singularity at the anomalous 
threshold for the weakly bound state as given by Eq.~(\ref{eq19-1}) in order to satisfy the DRs as shown in Fig.~\ref{fig4}.

Defining the intrinsic charge density $\rho ({\bf r})$ in three dimensional space, the transverse charge density $\rho(b)$ 
in two dimensional space and the longitudinal charge density $\rho_{\rm ILD} (r_z)$ in one dimensional space, respectively, 
in Eqs.~(\ref{eq23c}), (\ref{eq24c}) and (\ref{eq25c}), one may convince that the mean-square charge radius 
$\braket{r^2}_{\rm em}$ is given by the sum of the mean-square transverse radius $\braket{b^2}$ and the mean-square 
longitudinal distance $\braket{r^2_z}_{\rm ILD}$, i.e., $\braket{r^2}_{\rm em} = \braket{b^2} + \braket{r^2_z}_{\rm ILD}$.
While $\braket{r^2}_{\rm  ILD}$ and $\braket{r^2}_{\rm em}$ are derived from different spacetime dimensions and different methods,
both have the common factor $e_q m^2_{\bar Q} + e_{\bar Q} m^2_q$.
This observation allows to understand the negative value of mean-square charge radius of neutral mesons such as $K^{0}(d{\bar s})$ 
which have negatively charged light quark orbiting around the heavier ${\bar s}$ quark~\cite{Green}. 
The generic structures of longitudinal charge densities for charged particles ($\pi^+$, $K^+$, $D^+$) are similar to each other 
although the density profiles of charged particles are quite different from those of neutral ones ($K^0, D^0$).
For the case of equal constituent mass ($m_q = m_Q$), $\braket{r^2_z}_{\rm ILD}$ given by Eq.~(\ref{eq27d})
decreases monotonically to the minimum value $\braket{r^2_z}_{\rm ILD}\to 1/ (5 m^2_q)$ in the maximal binding limit
while $\braket{r^2_z}_{\rm ILD}\to\infty$ in the zero binding limit, which is consistent with the observation made in Ref.~\cite{GS90}. 
In contrast to the intrinsic charge density, the apparent charge density defined by the time component of the current depends on 
the reference frame and we note that $\braket{r^2_z }_{\rm BF}$ is smaller than $\braket{r^2_z }_{\rm ILD}$ due to the Lorentz contraction.

In terms of the newly introduced boost invariant variable $\tilde z$, we also define the longitudinal charged density in LFD 
as given by Eq.~(\ref{eqLF1}) noting that the LF longitudinal momentum fraction ${\bar \beta} = q^+/p^+$ is the variable conjugate 
to $\tilde z$.  
From Fig.~\ref{fig6}, we found that $\rho_{\rm LF}(\tilde{z})$ has a very long and oscillating tail behavior of $\tilde{z}$, consistent 
with the result shown in Ref.~\cite{MB19} for the case of two-constituents of a Fock-space component.

Comparing the results in $(1+1)$- and $(3+1)$-dimensions, we note that the effects of transverse momentum ${\bf k}_\perp$ reduce 
the charge radii and broaden the widths of the peaks in charge densities. 
While the qualitative behaviors of the form factors in both dimensions are not much different each other, their quantitative behaviors 
are quite sizable due to the effects of the transverse momenta of the quark and antiquark. 
We thus conclude that the transverse momentum plays the role of broadening the width of the resonance and 
significantly flattens the height of the corresponding form factor.
Our analysis of the solvable scalar field model can be extended to the phenomenologically more realistic LFQM 
as we have shown for the transition form factor $F_{\mathcal{M}\gamma}(q^2)$ in the meson-photon transition process, 
$\mathcal{M}(p) \to \gamma^*(q) + \gamma(p')$~\cite{CRJ17}.

\appendix*
\section{}
In this appendix we present the analytic forms of the valence and non-valence contributions to the form factor
in the spacelike region.
The explicit forms of $I^{q}_{\rm S1,S2}$ are
\begin{eqnarray}
\label{ap1}
I^{q}_{\rm S1}&=& \frac{g^2 C_S}{4\pi} \biggl[
c_{1} {\rm tan}^{-1}\left(\frac{\omega_Q}{\sqrt{1-\omega^2}}\right) + 
c_{2} {\rm tan}^{-1}\left(\frac{\omega_q}{\sqrt{1-\omega^2}}\right) 
\nonumber\\
&&\hspace{1cm} + c_{3}  {\rm tan}^{-1}\left(\frac{M^2(\beta -2) - (m^2_q - m^2_Q)\beta}{2 m_q m_Q \beta \sqrt{1-\omega^2}}\right)
\nonumber\\
&&\hspace{1cm} + c_{4} \ln \left(\frac{m^2_q \beta^2}{ \beta (m^2_q + {\bar\beta} m^2_Q) -M^2{\bar\beta}}\right)
\biggr],
\\
\label{ap2}
I^{q}_{\rm S2} &=& \frac{g^2C_S}{4\pi} \biggl[
d_{1} {\rm tanh}^{-1}\left(\frac{\sqrt{\gamma_Q}}{\sqrt{1+ \gamma_Q}}\right) + 
d_{2} {\rm tan}^{-1}\left(\frac{\omega_q}{\sqrt{1-\omega^2}}\right)
\nonumber\\
&&\hspace{1cm}
+ d_{3}  {\rm tan}^{-1}\left(\frac{M^2(\beta -2) - (m^2_q - m^2_Q)\beta}{2 m_q m_Q \beta \sqrt{1-\omega^2}}\right)
\nonumber\\
&&\hspace{1cm} + d_{4} \ln \left(\frac{m^2_q \beta^2}{ \beta (m^2_q + {\bar\beta} m^2_Q) -M^2{\bar\beta}}\right)
\biggr],
\end{eqnarray}
where 
\begin{equation}
C_S =\frac{1 -\omega^2}{8M^2 m^3_q m^3_Q (1 -\omega^2 + \gamma_Q)(1-\omega^2)^{3/2} (\beta^2-1)},
\end{equation}
and 
\bea
c_{1} &=& 4 m_q m_Q \omega \left( 2 M^2 {\bar\beta} + \beta Q^2 \right),
\nonumber\\
c_{2} &=& 8 m^2_q m^2_Q (1-\omega^2) + 2 {\bar\beta} [M^2 (m^2_q + m^2_Q) - (m^2_q - m^2_Q)^2]
\nonumber\\
&& + (M^2 - m^2_q + m^2_Q) Q^2 \beta,
 \nonumber\\
c_{3} &=& -4  m_q m_Q M^2 \omega {\bar\beta} 
\nonumber\\
&&-2  m_Q 
\left[4 m^2_q  m_Q \left(\omega^2-1\right) + m_q \omega Q^2 
-m_Q Q^2\right] \beta,
\nonumber\\
c_{4}&=& m_q m_Q \sqrt{1-\omega^2} \left(2 M^2+2 m^2_q -2 m^2_Q + Q^2 \right)\beta,
\eea
and
\bea
d_{1} &=& 4 M^2 m_q m_Q \left(\beta ^2-1\right) \sqrt{1-\omega^2}\frac{\sqrt{1+\gamma_Q}}{\sqrt{\gamma_Q}},
\nonumber\\
d_{2} &=& 
2 m_Q [ 
\left(\beta ^2-1\right) M^2 m_q \omega - {\bar\beta}^2 M^2 m_Q - 4 \beta  m^2_q m_Q (1 - \omega^2)
],
\nonumber\\
d_{3} &=& - c_{3}, 
\nonumber \\
d_{4} &=& - c_{4}.
\eea
The corresponding results in the timelike region are readily obtained by changing $Q^2\to -Q^2$ 
in Eqs.~(\ref{ap1}) and (\ref{ap2}).

\acknowledgments
H.-M.C. was supported by the National Research Foundation of Korea (NRF) under Grant No. NRF- 2020R1F1A1067990. 
The work of Y.C. and Y.O. was supported by NRF under Grants No. NRF-2020R1A2C1007597 and 
No. NRF-2018R1A6A1A06024970 (Basic Science Research Program).
C.-R.J. was supported in part by the US Department of Energy (Grant No. DE-FG02-03ER41260).

\end{document}